\documentclass[]{aa}
\usepackage{graphicx}
\usepackage{natbib}
\bibpunct{(}{)}{;}{a}{}{;}
\begin{document}
\title{Comparing Geometrical and Delay Radio Emission Heights in Pulsars}
\author{Dipanjan Mitra 
       \inst{1} 
	\and 
        X. H. Li
	\inst{1,2}}
	\institute{Max-Planck Institute f\"{u}r
	Radioastronomie, Auf dem H\"ugel 69, D-53121, Bonn, Germany
	\and National Astronomical Observatory, Chinese Academy of Sciences, 
        Jia-20 DaTun Road, Chaoyang District, Beijing 100012, China}

\offprints{D. Mitra, email: dmitra@mpifr-bonn.mpg.de}

\authorrunning{Mitra \& Li}
\titlerunning{Radio Emission Heights in Pulsars...}

\abstract{ We use a set of carefully selected published average
multifrequency polarimetric observations for six bright cone dominated
pulsars and devise a method to combine the multifrequency polarization
position angle (PPA) sweep traverses. We demonstrate that the
PPA traverse is in excellent agreement with the rotating vector model
over this broad frequency range confirming that radio emission
emanates from perfectly dipolar field lines. Correcting for the
effect of retardation we firmly establish the steepest gradient 
point in the combined PPA traverse to be the fiducial phase in these pulsars.
We use this combined
curve and inputs from earlier studies to determine the geometrical
angles of the neutron star namely $\alpha-$ the angle between the
rotation and the dipole magnetic axis and $\beta-$ the angle between
the magnetic axis and the observers line of sight. Using these
estimates of $\alpha$ and $\beta$ we derive the geometrical emission
heights ($\rm r_{\rm geo}$). Further using the 
relativistic beaming model based on effects of aberration and retardation
we find the delay emission heights 
$\rm r_{delay}^{\rm BCW}$ suggested by Blaskiewicz et al. (1991).
We find in general $\rm r_{delay}^{\rm BCW} < \rm r_{\rm geo}$
which can be explained by a broad emission region operating in pulsars 
or/and signature of magnetic field sweepback effect as suggested
by Dyks et al. (2003).

For pulsars with central core emission in our sample, we find 
the peak of central core component to lag the steepest gradient
of the PPA traverse at several frequencies. Also significant frequency
evolution of the core width is observed over this frequency range.
The above facts strongly suggest:
(a) the peak core emission does not lie on the fiducial
plane containing the dipole magnetic axis and the rotation axis, and (b) the core
emission does not originate from the polar cap surface.
 
}

\maketitle

\section{Introduction}
\label{sec1}

Pulsars exhibit broadband radio emission ranging over tens of MHz to
tens of GHz and the emission is highly polarized.  Inspection of pulse
shapes at multiple frequencies suggests that in general the pulse
radio emission beam consists of nested cones of emission, and a
central core emission (Mitra \& Deshpande 1999).  It is most often
seen that pulse widths at low observevable radio frequencies are wider
than that at higher frequencies. This observation, commonly referred
to as radius to frequency mapping (RFM) has been interpreted as low
frequency emission arising at higher altitude from the surface of the
neutron star while higher frequencies arise closer to the stellar
surface (Cordes 1978).

Radio emission from pulsars are thought to arise from dipolar magnetic
field lines. Evidence for this underlying magnetic field comes from
the observed 'S-shaped' swing of the polarization position angle (PPA)
across the pulse. To interpret this observed swing Radhakrishnan \&
Cooke (1969) proposed the rotating vector model (RVM) which is
interpreted in terms of radiation being beamed along the open dipolar
field lines and the plane of linear polarization is that containing
the field line associated the emission received at a given instant.
According to the model the PPA $\psi$ across the pulse phase $\phi$
can be represented as
\begin{equation}
\psi=\psi_{0}+\arctan\left(\frac{\sin\alpha \sin(\phi-\phi_{0})}{\sin\xi
 \cos\alpha-\sin\alpha \cos\xi\cos(\phi-\phi_{0})}\right).
 \label{eq1}
\end{equation}
(Manchester \& Taylor 1977).
Here $\alpha$ is the angle between the rotation axis and the magnetic
axis and $\beta$ is the angle between the rotation axis and the line
of sight to the pulsar.  $\psi_{0}$ and $\phi_{0}$ are the PPA and
phase offset of the observed curve respectively.  The angle $\xi =
\alpha + \beta$ is the angle between the observer's line of sight and
the pulsar rotation axis. While the RVM model has been successfully
applied to a large number of pulsars for several of them the PPA
traverse seem to show significant distortions. However even for
pulsars where the PPA traverses are in agreement with the RVM accurate
determination of $\alpha$ and $\beta$ is extremely difficult (Narayan
\& Vivekanand 1982, von Hoensbroech \& Xilouris 1997 HX97a hereafter).
Rankin (1990) suggested another method to estimate $\alpha$ for core
dominated pulsars.  She noticed that the 3 dB core widths at 1 GHz is
related to $\alpha$ and the pulsar period $\rm P$ as $2.45^{\circ}
\rm P^{-0.5}/\sin\alpha$.  This angular width corresponds roughly to the size of
the polar cap and thus implies that the core emission should originate from the
polar cap.  Further the RVM model can be used to calculate $\beta$.
However there has been suggestion that the core emission might
originate far from the stellar surface (Gil 1991), thus making it
conceptually difficult to use this method.

Accepting that the geometrical angles of the star can be determined
then by measuring the pulse width, the geometrical emission height
$\rm{r_{geo}}$ can be found (see Mitra \& Rankin 2002 hereafter MR02
for a recent review). Crucial thing to note here is that pulse width
measurements are based on two arbitrarily defined points on the
leading and trailing edge of the profile. And it is currently
impossible to determine the field line with respect to the last open
field line for these defined points. This uncertainty causes
difficultly in our understanding of what emission height
$\rm{r_{geo}}$ might be corresponding to.  We will come back to this
important question in a later section of this paper.

The RVM model was further advanced by Blaskiewicz et al. (1991,
hereafter BCW) where they included first order special relativistic
effects into account. The model predicts that due to aberration and
retardation (A/R) effects
the steepest gradient point of the PPA traverse has a time lag with respect to 
the center of the total intensity pulse profile.
This time lag can be converted to an emission height
$\rm{r_{delay}^{BCW}}$. Indeed observation confirms that for 95 \% of
the cases subject to such study the predicted lag is seen (BCW,
HX97a).  Hibschman \& Arons (2001) has redone the analysis of A/R
effects and confirmed the BCW results. Moreover the time lag seems to
increase with decreasing frequency in the way as predicted by RFM
(BCW, HX97a and Malov \& Suleimanova 1998).  
This method of estimating the emission heights has the
advantage that it is independent of the geometrical parameters
$\alpha$ and $\beta$. However this technique cannot be reliably
applied to several pulsars showing distorted PPA traverse where one
encounters difficulties in establishing the steepest gradient point in
the curve (a detailed discussion on this can be found in HX97a).

Recently, in a series of paper by Gangadhara \& Gupta (2001) and Gupta
\& Gangadhara (2003) (GGa and GGb hereafter) has applied the A/R
theory to total intensity profiles in pulsars and estimated emission
heights $\rm{r_{delay}^{GGa}}$. Although the formula derived by GGa
depend on $\alpha$ and $\beta$, recently Dyks
et. al. (2003) revisited the problem and have shown that
$\rm{r_{delay}^{GGa}}$ estimates are also independent of the
geometrical angles, just like the BCW method.

While there is no clear consensus, the above mentioned methods seem to
support the fact that RFM exists in pulsars.  Theoretical models for
pulsar emission predicts no RFM where the spectral index $a$ is 0
(Barnard \& Arons 1986) to a situation where $a$ is $0.66$ (Ruderman
\& Sutherland 1975). Also observationally this wide variety is seen
(Kijak \& Gil 1997, MR02, HX97a).  However it is often seen that the
estimated heights ${\rm r_{ geo}}$ and ${\rm r_{delay}}$ are in
disagreement (BCW). Comparison of these height estimates are often
difficult due to the enormous uncertainties involved in estimating
$\alpha$ and $\beta$. In this work, we use a set of six strong conal pulsars
to do a systematic analysis of RFM in detail. The selected data set
consists of multifrequency polarimetric data from Gould \& Lyne (1998)
and von Hoensbroech \& Xilouris 1997 (HX97b) and is discussed in
section~\ref{sec2}. We devise a method to combine PPA traverse at
multiple frequencies and find the corresponding $\alpha$ and $\beta$
as discussed in section~\ref{sec3}.  In section~\ref{sec4} we investigate
the inter-relationship between the improved ${\rm r_{ geo}}$ and ${\rm
r_{ delay}}$. In section~\ref{sec5} we summarize our results.

\section{Data selection}
\label{sec2}

Table~\ref{tab1} lists the pulsars we use for our analysis.  Pulsars
chosen is based on the criteria: (1) they have smoothly varying PPA
traverse observed over a frequency range of 0.4 -- 4.85 GHz.  and (2)
all the pulsars show clear evidence of outer conal components over
this whole frequency range. Using this criteria we ended up selecting
six strong pulsars for which multifrequency polarimetric data was
available from the European Pulsar Network (EPN) archive maintained by
Max-Planck Institut f\"ur Radioastronomie, Bonn\footnote
{http://www.mpifr-bonn.mpg.de/div/pulsar/data/}.  We used published
data at 0.4, 0.6, 0.9, 1.4 and 1.6 GHz from Gould \& Lyne (1998) and
at 4.85 GHz from HX97b.  In total we use 31 profiles spaced over this
frequency range.  The sample of pulsars belong to the morphological class
of conal double (D), triple (T) and multiple (M) as classified by
Rankin (1993) and mentioned in table~\ref{tab1}.

Three pulsars in our sample namely PSR B1737+17, PSR B2045$-$16 and
PSR B2111+46 have clearly identifiable core components.
Classification of the core component is based on the sign changing
circular polarization observed in average profiles as suggested by
Rankin (1990).  A sign changing circular is also seen for PSR
B1039$-$19 in the leading part of the profile, however as suggested by
Han et. al. (1998) in several complex profiles such sign changing
circular is not only restricted to the core component. Based on the
location of this component, it is more demanding that this feature is
not associated with the central core emission. Further multifrequency
Gaussian fits (see Kramer 1994) to the profiles shows that 4 or 5 components fits the
pulse shape well for this pulsar, thus making PSR B1039$-$19 more a M
type pulsar.

The full Stokes ($I,Q,U,V$) parameter data is used to construct the
PPA, $\psi = 0.5 \tan^{-1}(U/Q)$. The linear polarization is first
calculated as $\sqrt{U^2 + Q^2}$ and then the off-pulse mean is
subtracted to remove the positive bias. PPA points for which the
linear polarization exceeds 3 times the off-pulse noise of the linear
polarization is used in our analysis.  The error on $\psi$ is
calculated as $\sigma_{\psi} = \sqrt{ (Q \sigma_Q)^2 + (U
\sigma_{U})^2}/2L^2$ , where $\sigma_Q$ and $\sigma_U$ are the off
pulse rms in Stokes $Q$ and $U$ respectively. It has been recently
shown by Li \& Han (2003) that the PPA traverse can be affected
strongly by interstellar scattering.  We note that none of the pulsars
in our sample show any scattering, and thus the PPA traverse mostly
reflect intrinsic properties within the pulsar magnetosphere.

\begin{table*}
\caption{The table below lists the pulsar parameter used in our
analysis. Columns 1, 2 and 3 gives the PSR Bname, Period and the
morphological type classified by Rankin (1993a). Columns 4, 5 and 6
gives the values of $\alpha$, $\beta$ and reduced $\chi^2$ obtained by
us (indicated as our) after the combined fits. Columns 7, 8, 9 and 10
give the $\alpha$ and $\beta$ from Lyne \& Manchester (1988, indicated
as LM88) and Rankin (1993b, indicated as R93). The superscript dagger
corresponding to these $\alpha$ values are cases where one needs to do
the transformation (180 - $\alpha$) to compare with our values.
Columns 11 gives the value of the opening angle of the polar cap
$\rho_{\rm pc}$.
\label{tab1}}
\centering
\begin{tabular}{cccr@{$\pm$}lr@{$\pm$}lrrrrrr@{$\pm$}l}
\noalign{\smallskip} 
\hline 
\hline
\noalign{\smallskip} 
PSR Bname & Period (s) & CLASS & \multicolumn{2}{c}{$\alpha\,(^{\circ})$} &\multicolumn{2}{c}{$\beta\,(^{\circ})$} & $\chi^2$ &$\alpha\,(^{\circ})$&$\beta\,(^{\circ})$&$\alpha\,(^{\circ})$&$\beta\,(^{\circ})$ &\multicolumn{2}{c}{$\rho_{\rm pc}$}\\
&  &  & \multicolumn{2}{c}{(our)} &\multicolumn{2}{c}{(our)} & & R93& R93&LM88&LM88 &\multicolumn{2}{c}{($^{\circ}$)}\\
\noalign{\smallskip} 
\hline 
\noalign{\smallskip} 
B0301+19   & 1.387 & D  &159&16 & $-$1.1&0.5&2.6  &31.9$^{\dagger}$&1.8&30$^{\dagger}$&1.7 & 1.05 & 0.3  \\
B0525+21   & 3.745 & D  &54 &4  & 1.3   &0.1&4.1  &23.2&0.7&21&0.6& 0.64&0.2 \\
B1039$-$19 & 1.386&  M  &51 &19 &    2.5&0.3&2.5  &33.8&1.8&31&1.7& 1.05&0.3 \\
B1737+13   & 0.803 & M  &47 &12 &    2.1&0.4&5.6  &40.8&2.5&41&1.9& 1.38&0.4 \\
B2045$-$16 & 1.961 & T  &127&2  & $-$1.4&0.1&12.1 &36.7$^{\dagger}$&1.1&36$^{\dagger}$&1.1&0.88&0.3 \\
B2111+46   & 1.014 & T  &14 &4  & $-$1.4&0.5&21.3 & 8.6&1.3& 9&1.4& 1.23&0.4 \\
\noalign{\smallskip} 
\hline 
\hline
\end{tabular}
\end{table*}

\section{Determining the viewing geometry}
\label{sec3}

\begin{figure*}
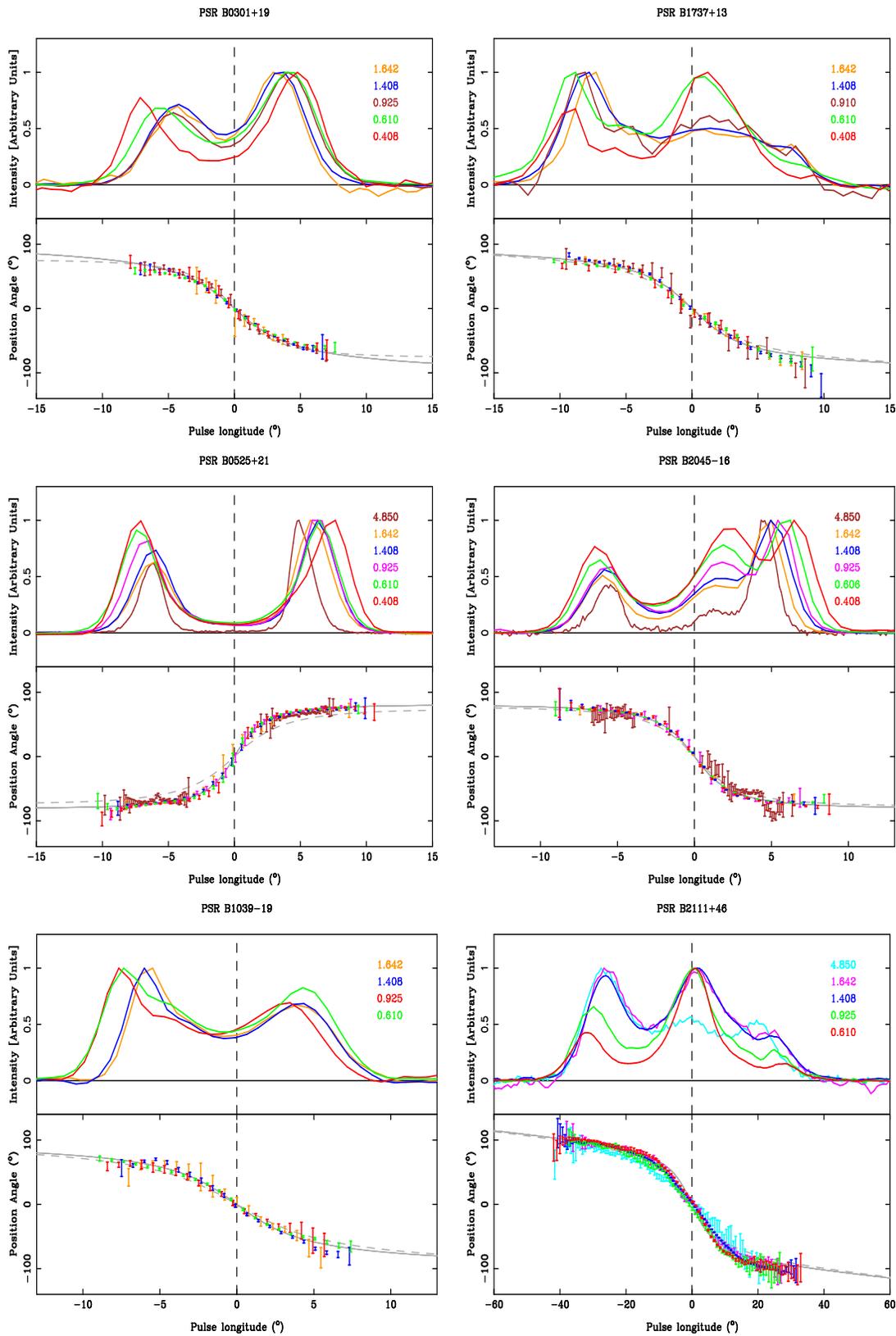

\begin{tabular}{@{}lr@{}}
{\mbox{\includegraphics[width=7cm,height=7cm,angle=-90]{B0301_all.cps}}}&
{\mbox{\includegraphics[width=7cm,height=7cm,angle=-90]{B1737_all.cps}}}\\
{\mbox{\includegraphics[width=7cm,height=7cm,angle=-90]{B0525_all.cps}}}&
{\mbox{\includegraphics[width=7cm,height=7cm,angle=-90]{B2045_all.cps}}}\\
{\mbox{\includegraphics[width=7cm,height=7cm,angle=-90]{B1039_all.cps}}}&
{\mbox{\includegraphics[width=7cm,height=7cm,angle=-90]{B2111_all.cps}}}\\
\end{tabular}
\caption{The above set of plots shows the profiles aligned with
respect to the center of the PPA traverse at various frequencies.  In
the top panel the total intensity of the profiles are plotted by lines
with different colours. The colour coding used for every frequency is
indicated in the top panel. The bottom panel show the combined PPA
traverse and the same symbol coding is used. The central x=0 line is
indicated by the dashed line in both panel. The solid gray line is the
best fit RVM to the combined PPA traverse and the dashed line corresponds
to RVM using values of the viewing geometry from Rankin (1993).}
\label{fig1}
\end{figure*}

If RVM is assumed, then the PPA traverse observed at several
frequencies should be the same as the viewing geometry is constant for
a given star (see equation~\ref{eq1}). The only difference is that with
decreasing frequency the overall broadening of the PPA traverse will
extend over wider longitude ranges.  Thus in principal fitting RVM
model to the lowest observable frequency should give the best solution
for the viewing geometry. On the other hand for most of the conal
double profiles (e.g. PSR B0525+21) the linear polarization towards
the center of the profile is low and thus the PPA has large
uncertainties which is more prominently seen at lower frequencies.
Thus to fill up the PPA traverse for these central longitudes the high
frequency profiles are helpful.  Hence it is desirable to combine all
the available frequencies and fit the RVM to the combined PPA traverse
to obtain better estimates of $\alpha$ and $\beta$.  It is to be noted
that the pulse widths needs to be as large as $\sim 50^{\circ}$ in
order to be able to improve estimates of $\alpha$ and $\beta$
significantly due to this reason.

According to the BCW model, the behaviour of the PPA traverse depends
on the emission height and the pulsar period (see equation~16 in BCW). This
A/R effect is enhanced in millisecond pulsars (Gil \& Krawczyk 1997)
where the pulsar rotates significantly faster. While for normal slower
pulsars as in our sample BCW already notes that a phase shifted RVM
fits the PPA traverse reasonably well. A similar conclusion has also
been reached by Hibschman \& Arons (2001). Considering this to be the
premise for interpreting the multifrequency PPA traverse, in the rest
of the section we outline the ways by which we are able to combine
the multifrequency PPA traverse and estimate the viewing geometry.

\subsection{Earlier studies dealing with viewing geometry}

The prime difficulty encountered while fitting the RVM given by
equation~\ref{eq1} to the PPA swing curve is that several combination of
$\alpha$ and $\beta$ can produce equally acceptable fits (see HX97a).
Further Narayan \& Vivekanand (1982) pointed out that to constrain
these angles better one needs the PPA traverse to extend over wide
longitude ranges and particularly that the only way to discriminate if
the line of sight passes outside (outer) or inside (inner) the
rotation axis and the magnetic axis is to have wider profiles.  A
detailed description of RVM fitting proccedure and definitions can be
found in Everett \& Weisberg (2001, hereafter EW).

This work of EW reviews the various methods that has been used to
obtain the viewing geometry in the past and further fits RVM to
several pulsars using a rigorous treatment of statistical errors on
PPA's. In order to include sufficient number of points in the profile
wings they considered points for which signal to noise ratio (SNR) of
the unbiased linear polarization exceeds 1.57. This is however
somewhat confusing and unphysical as they retain points even at
longitude ranges where there is no pulsed emission i.e. beyond the
open dipolar field lines. On comparison, we have considered points
with SNR greater than 3, and thus are certain that lower SNR points
are not affecting the fits.

Nonetheless the covariant nature of $\alpha$ and $\beta$ makes it
impossible to constrain the viewing geometry. The study of Lyne and
Manchester (1988) and Rankin (1990, 1993) here is of great importance,
as they have tried to use the profile width information (which is more
sensitive to $\alpha$ than $\beta$) along with the steepest gradient
of the PPA traverse to constrain the viewing geometry (see EW for
detailed description). The remarkable aspect of the above two studies
is that from two different set of arguments they obtain very similar
values of $\alpha$ and $\beta$. Hence we have used their values as
initial guesses while fitting the PPA traverse. Thus here
we have two major goals: (a) To examine the validity of RVM over a wide
frequency range, (b) To assess if RFM can obtain wider PPAs near the
wings of the profile enabling better estimation of the
viewing geometry.

\subsection{The problem of PPA moding}

EW noticed that several average RVM-style PPA traverse
on close inspection tends to be slightly incorrect. A possible reason
for this is that pulsar radiation is often in the form of two
orthogonal modes, and mixing of these modes at some longitude ranges
in the pulse can corrupt the average classical RVM. Such orthogonal as
well as non-orthogonal moding has been observed in several
pulsars. There is a danger thus in using average PPA traverse to
obtain any pulsar parameters.  Here as we will see later, most of
the pulsars considered follow the RVM reasonably well.  There are two
pulsars in our sample PSR B0301+19 and B0525+21 where comparison of
single pulse and average profiles are available (Backer \& Rankin
1980, Stinebring et al. 1984). For both cases the average PPA is in
good agreement with the single pulse PPA. Only towards the wings of
the profiles some low level modal effects are seen (see Rankin \&
Ramachandran 2003 for discussion).  Strong modal effects should
reflect on phase dependent distortion of the PPA traverse, and are
known to vary with frequency (Karastergiou et al. 2002).  Our sample
pulsars however are free from such distortions.

\subsection{Combining the multifrequency PPA traverse}

As a first step we fit the RVM given by equation~\ref{eq1} to a given
pulsar at every available frequency.  The fits are done using
primarily the method outlined in HX97a to obtain the best fit values
for $\alpha$, $\beta$, $\phi_{0}$ and $\psi_{0}$.  The method uses the
Levenberg-Marquart Algorithm (Marquart 1963) as implemented in
Numerical recipes (Press et al. 1986) and is sensitive to the initial
parameters.  As initial inputs we have used $\alpha$ and $\beta$
values from Rankin (1993). The values obtained by Rankin (1993) and
Lyne \& Manchester (1988) are listed in table~\ref{tab1}, which as
seen are very similar. This unconstrained fit resulted in highly
correlated values of $\alpha$ and $\beta$.  The formal 1$\sigma$
errors in $\alpha$, $\beta$ $\psi_{0}$ and $\phi_{0}$ are obtained
from the $\chi^2$ fitting procedure using the resultant covariance
matrix as prescribed in Press et al. (1986). The phase of the steepest
gradient point $\phi_{0}$ in the RVM fit is found by zeroing the
second derivative of equation~\ref{eq1} with respect to the pulse phase
$\phi$ as discussed in HX97a. The values are found to be in good
agreement with that obtained from the formal fitting procedure.

In columns 3, 4 and 5 of table~\ref{tab2} the estimated $\alpha$,
$\beta$ and the reduced $\chi^2$ values for all pulsars at various
frequencies are given.  The steepest gradient of the PPA traverse
$(\partial \psi / \partial \phi)_{\rm max} = \sin\alpha/\sin\beta$ is
tabulated in column 6. 
At this stage we note that the RVM fits are in excellent agreement 
for each pulsar over this wide frequency range.
Thus we are now in a position to combine the PPA traverse. We do this
firstly by subtracting the offsets $\phi_{0}$ and $\psi_{0}$ of the
PPA traverse at each frequency and then overlapping them on each
other. The combined PPA traverse for all the pulsars are shown in
figure~\ref{fig1}. This combined PPA traverses are now fitted to obtain
$\alpha$ and $\beta$.  These values eventually obtained have
significantly smaller error bars than the individual fits. However
$\alpha$ and $\beta$ still remains highly correlated.  Note that the
reduced $\chi^2$ value for PSR B2045$-$16 and B2111+46 are somewhat
large, which is due to the 4.85 GHz PPA traverse showing larger
deviation from the other frequencies.  However for 5 of the cases the
profile widths are not wide enough to find if the lines of sight are
inner or outer.  Only for PSR B2111+46 the line of sight is inner and
consistent with the findings of Athanasiadis et al. (2003).  It is
noteworthy that we have obtained convergent fits for PSR B0301+19,
B0525+21 and B2045$-$16. For all these pulsars different inputs of
$\alpha$ and $\beta$ obtained from previous studies converged to the
same solution.

The smaller error estimates of $\alpha$ and $\beta$ are possible
primarily due to (a) large number of independent points available in
the combined PPA to fit the RVM and (b) RFM enabling parts of the PPA
traverse to be filled up with independent measurements. The estimated
error reflects the covariant relationship between these
parameters. Unfortunately there is little improvement between the nature of
correlation between $\alpha$ and $\beta$ for this combined PPA
traverse compared to the curves at individual frequency. For
comparison we show by dashed curve in the bottom panel of
figure \ref{fig1} for which $\alpha$ and $\beta$ are used from Rankin
(1993b). For most of the cases the solid and the dashed lines are
indistinguisable reflects the highly correlation between
$\alpha$ and $\beta$. Only for PSR B0525+21 our fitted line (solid)
is better than the literature value.  The reason for this high
correlation is most of the pulsars in our sample have widths of $\sim
20^{\circ}$, not being sensitive enough to affect the fits. And also
increase in width is only marginal, typically a factor of 1.3 between
the highest to lowest frequency as seen in column 7, 8 and 9 in
table~\ref{tab2}. Thus in general to obtain better estimates of the
viewing geometry low frequency polarimetric observations are
absolutely essential where due to RFM wider profiles are observed.

Finally the agreement of the combined PPA traverses with the 
RVM over this wide frequency range can be 
used to identify the fiducial phases for these pulsars. 
The fiducial phase is defined as the point lying in the plane containing the 
dipole magnetic axis, the rotation axis and the observers line of sight. 
The steepest gradient point in the PPA traverse is identified as such
a point according to RVM.
Due to the A/R effects proposed by the BCW model the steepest gradient point
is retarded towards later longitudes by an amount proportional to $\rm r(\nu)/c$
in a frequency ($\nu$) dependent manner, where $\rm r(\nu)$ is the emission height
and $\rm c$ is the velocity of light.
In the process of producing combined  PPA traverses we have eliminated the effect
of retardation. And given the excellent symmetry
of the combined PPA traverses observed for all the pulsars we conclude that
the inflexion or the steepest gradient point undoubtedly is the location
of fiducial phase in these pulsars. This fiducial phase should be used
for cold plasma dispersion correction and timing analysis in pulsars.

\section{Comparing radio emission heights}
\label{sec4}

There exists primarily two methods widely used to obtain radio
emission heights in pulsars. In this section we will first mention
briefly these methods and then subsequently compare the emission
heights for our sample of pulsars obtained using these methods.

\begin{figure*}
\begin{tabular}{lr@{}lr@{}lr@{}}
{\mbox{\includegraphics[width=5cm,height=5cm,angle=-90]{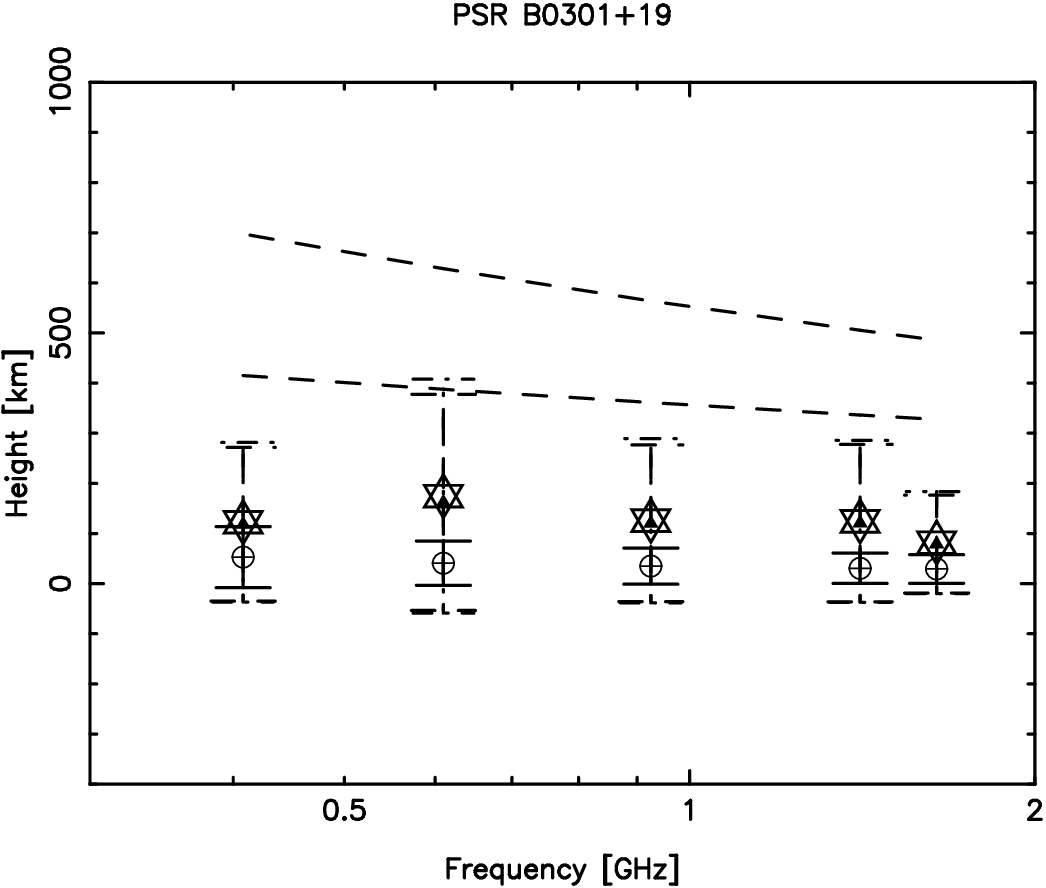}}}&
{\mbox{\includegraphics[width=5cm,height=5cm,angle=-90]{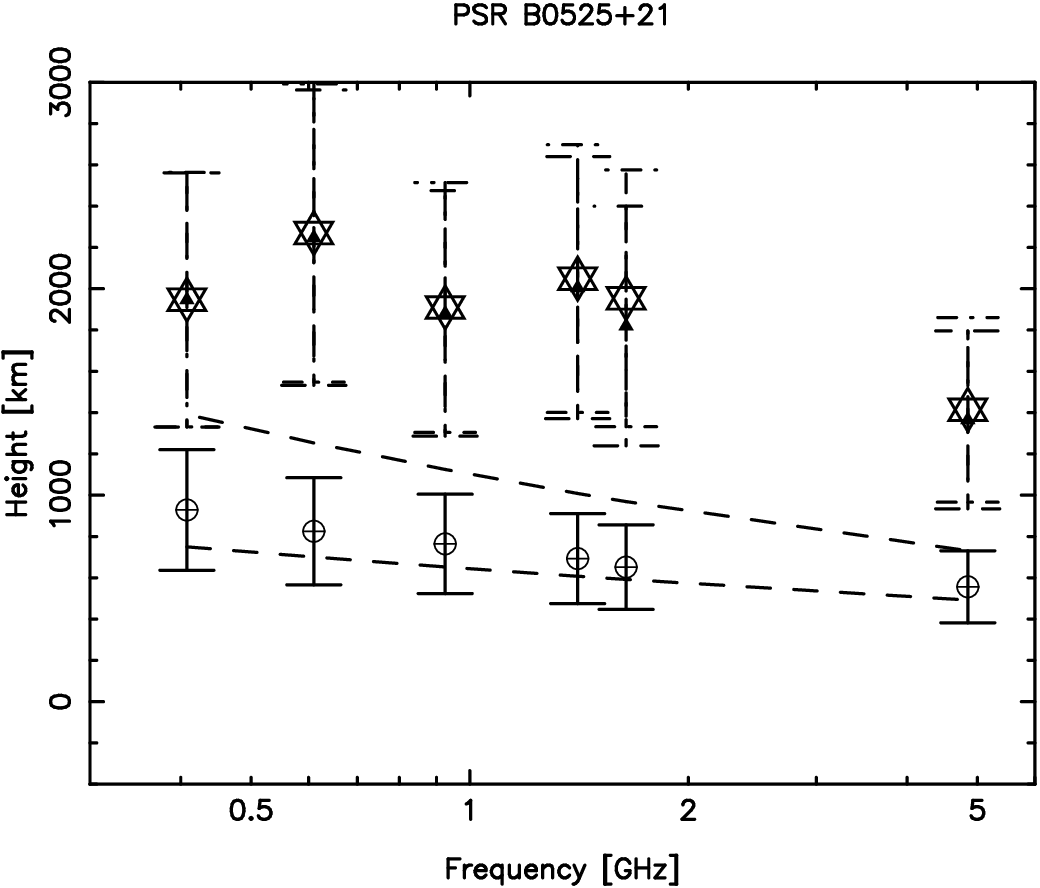}}}&
{\mbox{\includegraphics[width=5cm,height=5cm,angle=-90]{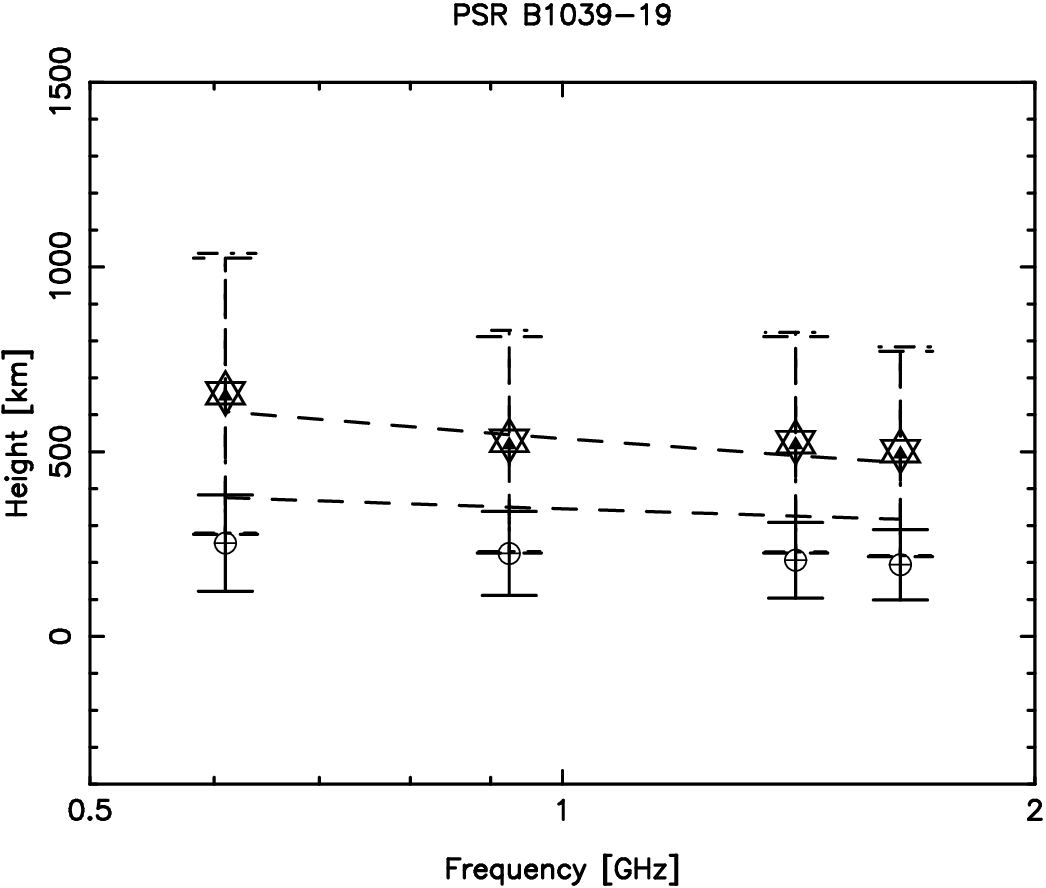}}}\\
{\mbox{\includegraphics[width=5cm,height=5cm,angle=-90]{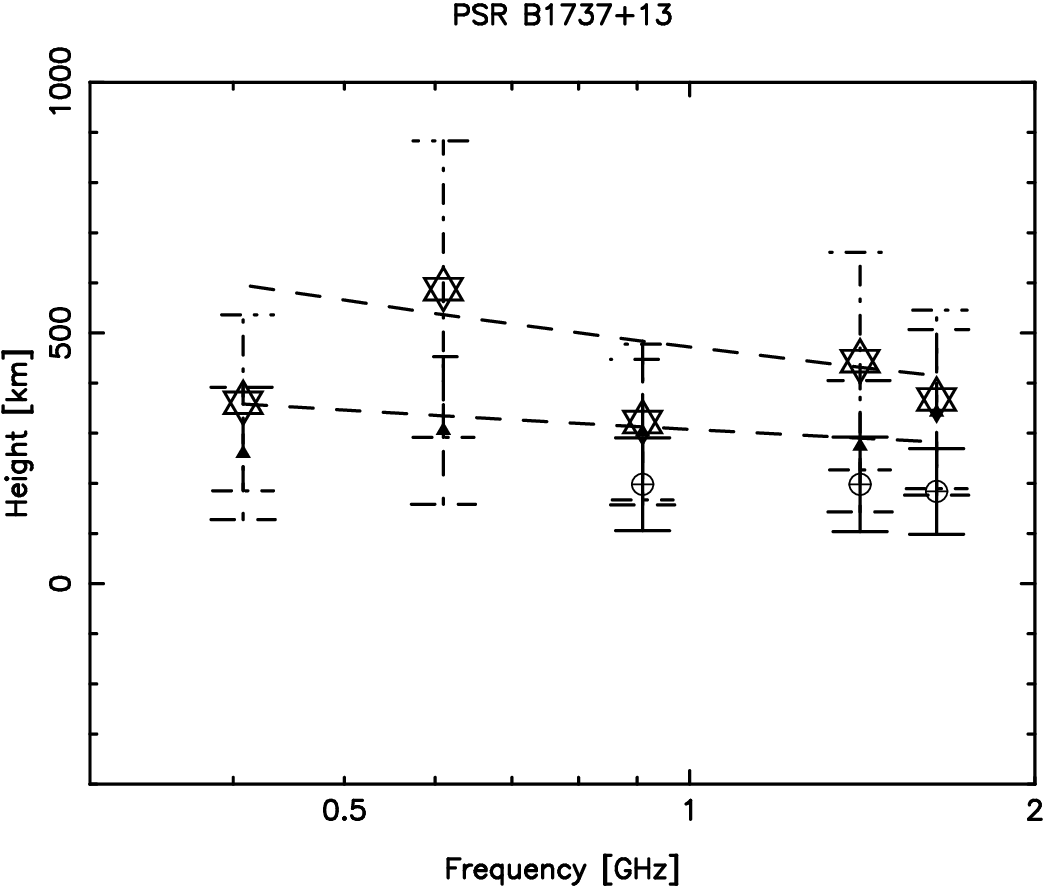}}}&
{\mbox{\includegraphics[width=5cm,height=5cm,angle=-90]{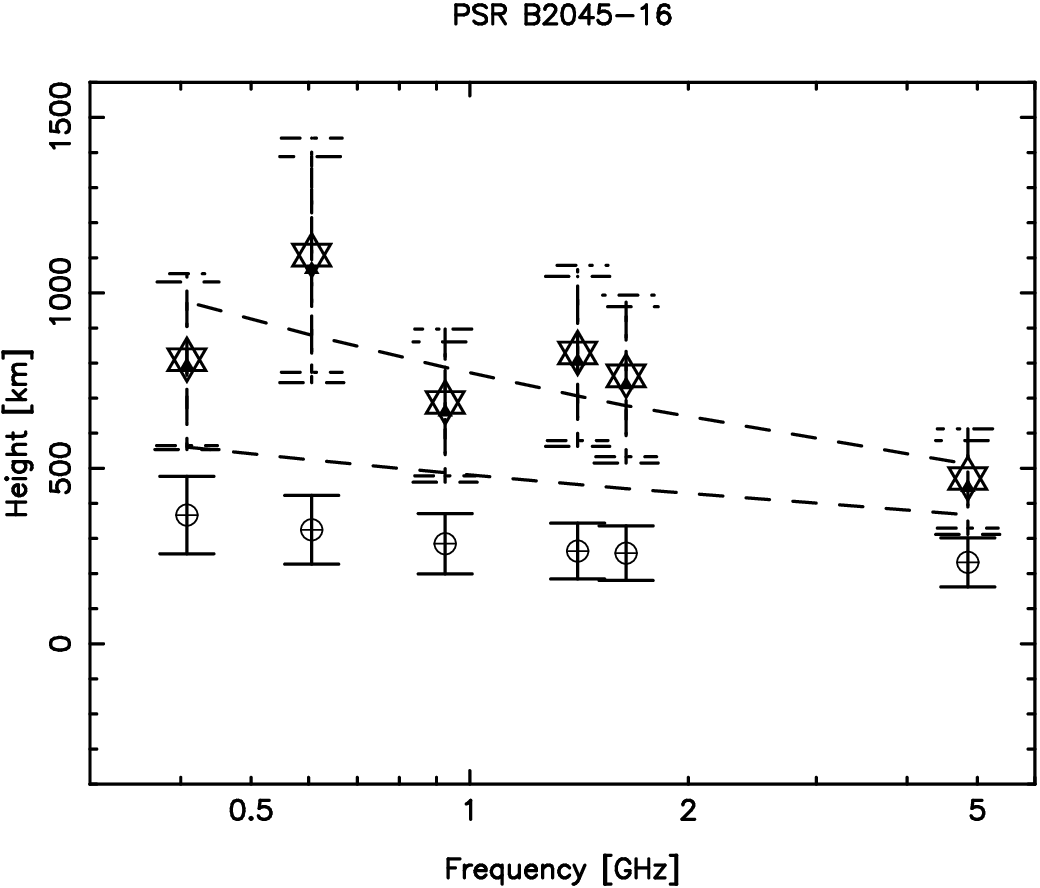}}}&
{\mbox{\includegraphics[width=5cm,height=5cm,angle=-90]{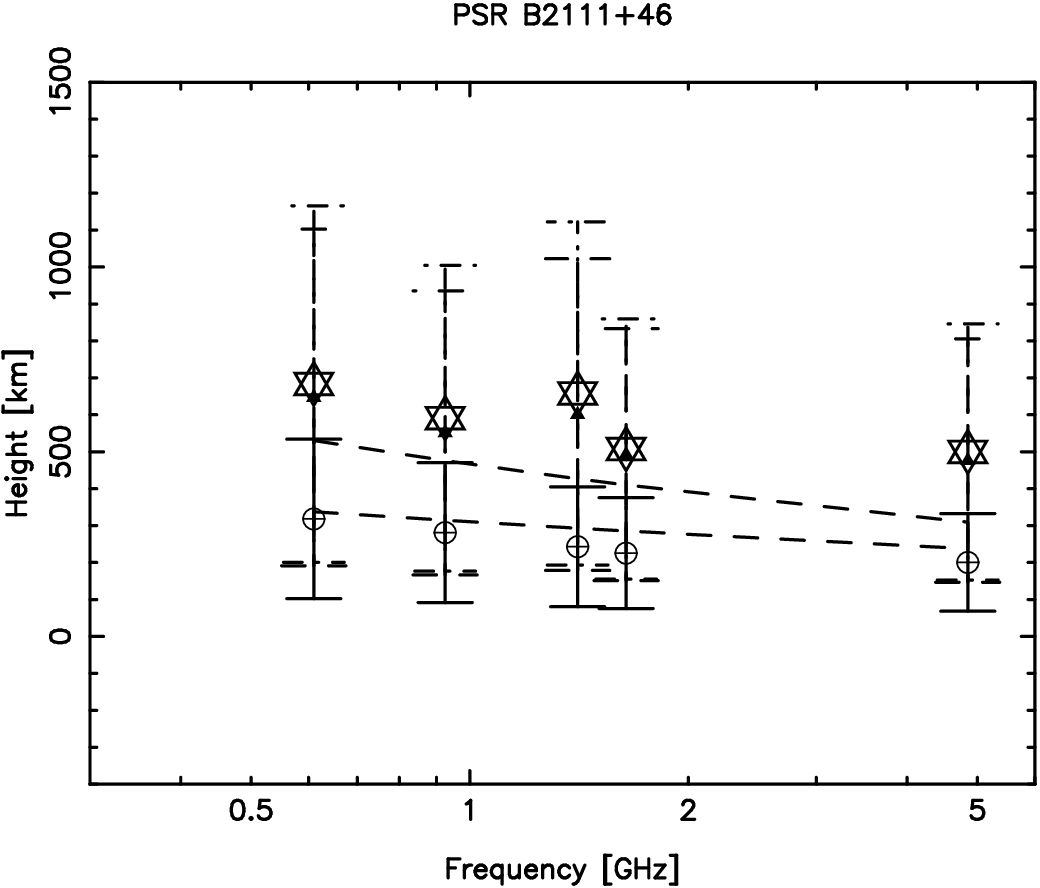}}}\\
\end{tabular}
\caption{In the figures the frequency dependence of $\rm r_{\rm geo}$ for
PPCS (circles with crosses), OTS (stars) and OTHX (filled triangles)
measurements are shown. Here $\rm r_{\rm geo}$ calculated assumes
$s=1$ for each case.  The dashed line correspond to upper and lower
limits of the $\rm r_{\rm delay}^{\rm KG}$ emission heights See text for further detail.}
\label{fig2}
\end{figure*}

\subsection{Emission Heigths using the Geometrical method: ${\rm r}_{\rm geo}$}

Assuming emission arising from open dipolar field lines and circular
emission beam shape, geometrical emission height ${\rm r_{geo}}$ can
be found.  The formula (see Kijak \& Gil 1997) is given by,
\begin{equation}
{\rm r_{geo}} \simeq \rm R_{*} {\rm P} \left(\frac{\rho}{1^{\circ}.23}\right)^2 s^{-2}~\rm{km.}
\label{eq2}
\end{equation}
Here $\rho$ is the radius of the emission beam calculated as
$\sin^2(\rho/2) = \sin(\alpha + \beta)\sin(\alpha)\sin^2(\rm{W}/4)
+\sin^2(\beta/2)$, (Gil, Gronkowski \& Rudnicki 1984).  ${\rm P}$ is
the period of the pulsar in seconds and $\rm{W}$ is the measured pulse
width. The parameter $s$ lying between $0 \leq s \leq 1$ 
describes the locus of the field line at 
the polar cap , where $s=0$ at the magnetic pole and $s=1$ at 
the edge of the Goldreich-Julian (1969) circular polar cap.
The radius $\rm R_{*}$ entering the equation is most often
assumed to be 10 km. Constrain for $\rm R_{*}$ from various equation of
states of neutron stars as well as simple blackbody fits to thermal
x-ray emission from pulsar PSR B0656+14 predicts a range of radius which can
vary by about 30\% (Brisken et al. 2003). In our subsequent
calculations of ${\rm r_{geo}}$ we include this spread of
$\rm R_{*}$ in the error estimates.  However before we estimate
$\rm{r_{geo}}$ for our sample of pulsars, we first review the factors
affecting determination of $\rm{r_{geo}}$.

The beam radius $\rho$ appearing in equation~\ref{eq2} requires accurate
knowledge of the viewing geometry $\alpha$, $\beta$ and the pulse
width $\rm{W}$. As we have discussed in the earlier section, even by
using the most exhaustive information of multifrequency PPA traverses,
exact determination of $\alpha$ and $\beta$ remains questionable due
to the covarient nature of these parameters.  Nonetheless if we accept
these values, then by measuring $\rm{W}$ based on two defined points
on the leading and the trailing edge of the profile $\rho$ and
consequently ${\rm r_{\rm geo}}$ can be found. ${\rm r_{\rm geo}}$
corresponds to the emission height of these two defined points
measured from the center of the neutron star provided we assume that
the beam is circular and the two defined points on either side of the
profile has the same emission height. The difficulty however arises in
identifying these defined points.  In the work of MR02 they noted that
for the outer conal component $\rm{W}$ measured using the outer peak
to peak component separation (PPCS) are far more well behaved with
frequency compared to outside half-power (3 dB) or outside 10\%
profile width. Mitra \& Deshpande (1999) showed that the pulsar beam
is consistant with a circular shape, however their analysis was based
only on PPCS measurements. It is important to note that $\rho$ of the
pulsar beam changes with frequency and the defined outer edge points
at different frequency corresponds to different dipolar field lines.
Thus a stable variation of $\rho$ (or $\rm{W}$ or $\rm r_{\rm geo}$) with
frequency implies that the locus of the defined intensity points on
these dipolar field lines follow a smoothly varying curve (not
necessarily a circle). If $\rho$ obtained from two sets of such
uniformly defined points (e.g. the peak to peak componets
separation and the 10\% widths) varies differently with frequency, it
only means that the intensity distribution of these two defined points
are not concentric.

To get further insight into this problem (and for other reasons
mentioned later on), here we measure profile widths based on three
different definition (1) The PPCS, where we fit each profile with
Gaussians to obtain the outer conal peaks accurately.  (2) The outer
three sigma (OTS) method, which involves choosing outer edges of the
pulse profile which are at three times the off pulse $rms$ level, and
(3) Using the outer three sigma method of HX97a (OTHX), where we first
choose the outer three times $rms$ point for the weaker component and
then calculate the percentage level at that point with respect to the
peak intensity of that component. Then we estimate the width by finding
the location of the same percentage level with respect to the peak for
the stronger component.  For cases (2) and (3) an interpolation within
the profile resolution was done to match the desired defined phases.
The error in the phase ($\phi$) for method (1) is obtained from the
formal Gaussian fits while for cases (2) and (3) we use the relation
suggested by HX97a namely,
\begin{equation}
\sigma_{\phi} = rms \frac{\tau}{\Delta I}
\label{eq4}
\end{equation}
where $\tau$ is the resolution of observation and $\Delta I$ is the
gradient of the total intensity around the defined phase points. The
measured widths and the errors are given in table~\ref{tab2}. For
PSR B1737+17 the trailing component is seen to only appear clearly at
frequencies 910, 1408 an 1642 MHz.  Thus PPCS widths are reliable in
these frequencies and only these values are quoted. Also the OTS and
OTHX points are better measured at these frequencies.

Having chosen the outer points the crucial difficulty is to know the
field line for these defined points with respect to the last open
field line, i.e the $s$ parameter. For outer PPCS measurements MR02
and GGb finds $s\sim0.6$. Kijak \& Gil (1997) tried to overcome this
problem by measuring $\rm {W}$ based on outer conal edge phase points
down to the lowest possible intensity level of about 1\% and using
$s=1$ for these points. They applied their method to a number of
pulsars and finds an empirical relation for emission height $\rm
{r_{\rm geo}^{\rm KG} \rm (km)} = (400\pm80)\nu_{\rm
GHz}^{-0.26\pm0.09}\dot{\rm P_{15}}^{0.07\pm0.03} \rm P^{0.3\pm0.05}$,
where $\dot{\rm P}$ is the period derivative in units of 10$^{-15}$ s/s of the pulsar (Kijak \&
Gil 2003). Since this relation is derived statistically using a large
sample of pulsars, the errors introduced by the questionable viewing
geometry is reduced. Nonetheless the unknown $s$ parameter does
introduce uncertainty in estimating $\rm{ r_{geo}}$.

In figure~\ref{fig2} we plot $\rm r_{\rm geo}$ obtained by three
different profile widths, and for comparison the upper and lower
limits $\rm r_{\rm geo}^{\rm KG}$ relation is plotted as dashed lines.
The following conclusions are apparent from figure~\ref{fig2}.

(a) Emission heights obtained using PPCS decrease with increasing
frequency in a rather stable manner.  The OTS, OTHX together with the
10\% and 3 dB widths noted by MR02 show more jittery behaviour with
frequency. However the OTS and OTHX points are consistent with a
stable variation within the error bars as seen figure~\ref{fig3}.  High
resolution and high quality signal to noise profiles are necessary to
investigate this feature in detail.

(b) For five cases in our sample the OTS and OTHX emission heights is
in good agreement with $\rm r_{\rm geo}^{\rm KG}$, given the error
bars. Hence $\rm r_{\rm geo}^{\rm KG}$ can be considered as robust
upper limits for the emission height of the outer edge of the pulsar
beam, although there might be deviations for individual cases.  Only
for the longest period pulsar PSR B0525+21 we find that for OTS and
OTHX measurements $\rm r_{\rm geo}> \rm r_{\rm geo}^{\rm KG}$.  This
can arise due to the fact that the empirical fit of $\rm r_{\rm
geo}^{\rm KG}$ at the long period ($>2$ sec) end is not well
constrained due to small number statistics.

Noticing that $\rm r_{\rm geo}$ measured using OTS and OTHX methods
are in agreement within the error bars, we proceed only considering
results corresponding to the OTS method since the results would be
similar if OTHX methods were used.

\subsection{Delay emission heights: $\rm r_{\rm delay}$}

Based on the kinematical effects ( A/R ) BCW and GGa devised two
independent ways to calculate delay emission heights $\rm r_{\rm
delay}$ as mentioned earlier.  What makes the delay methods attractive
for measuring emission heights is that it is independent of the
viewing geometry. Both the methods assumes that the pulsar beam is 
symmetric about the fiducial phase.
In this section we will discuss these two methods
in succession and apply them to our sample of pulsars wherever possible.

\begin{figure*}
\begin{tabular}{lr@{}lr@{}lr@{}}
{\mbox{\includegraphics[width=5cm,height=5cm,angle=-90]{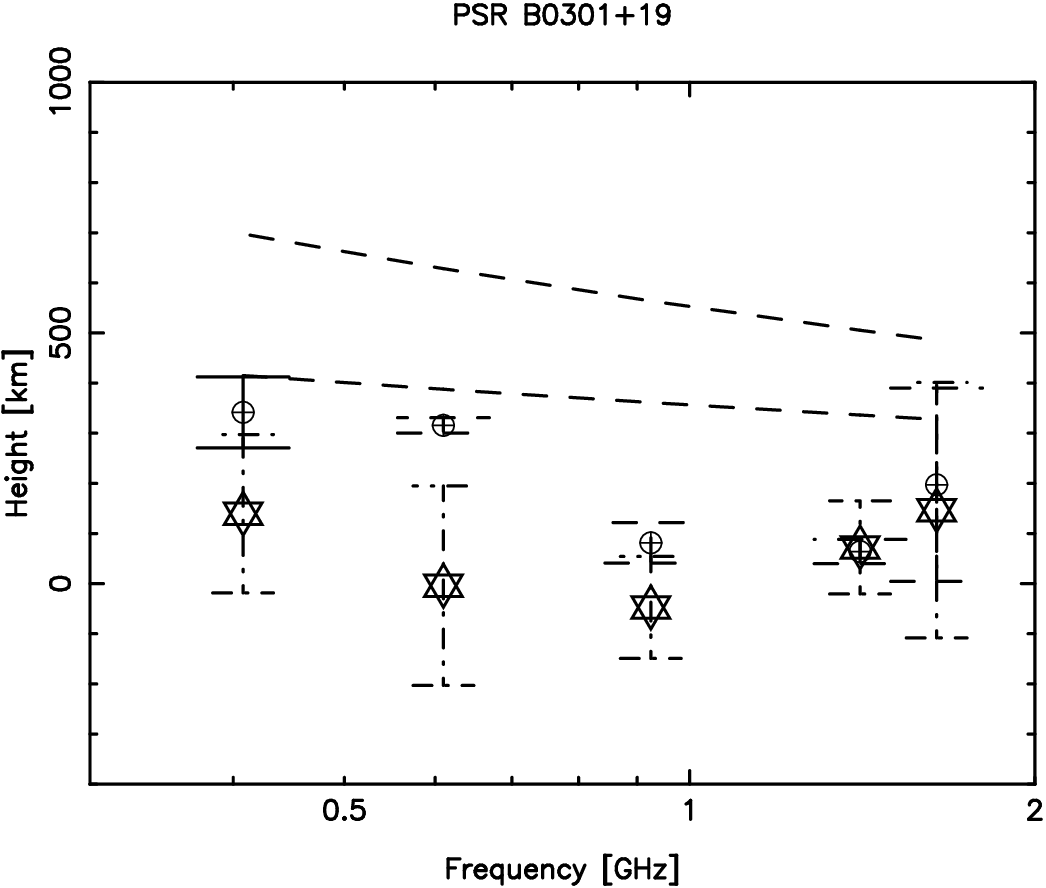}}}&
{\mbox{\includegraphics[width=5cm,height=5cm,angle=-90]{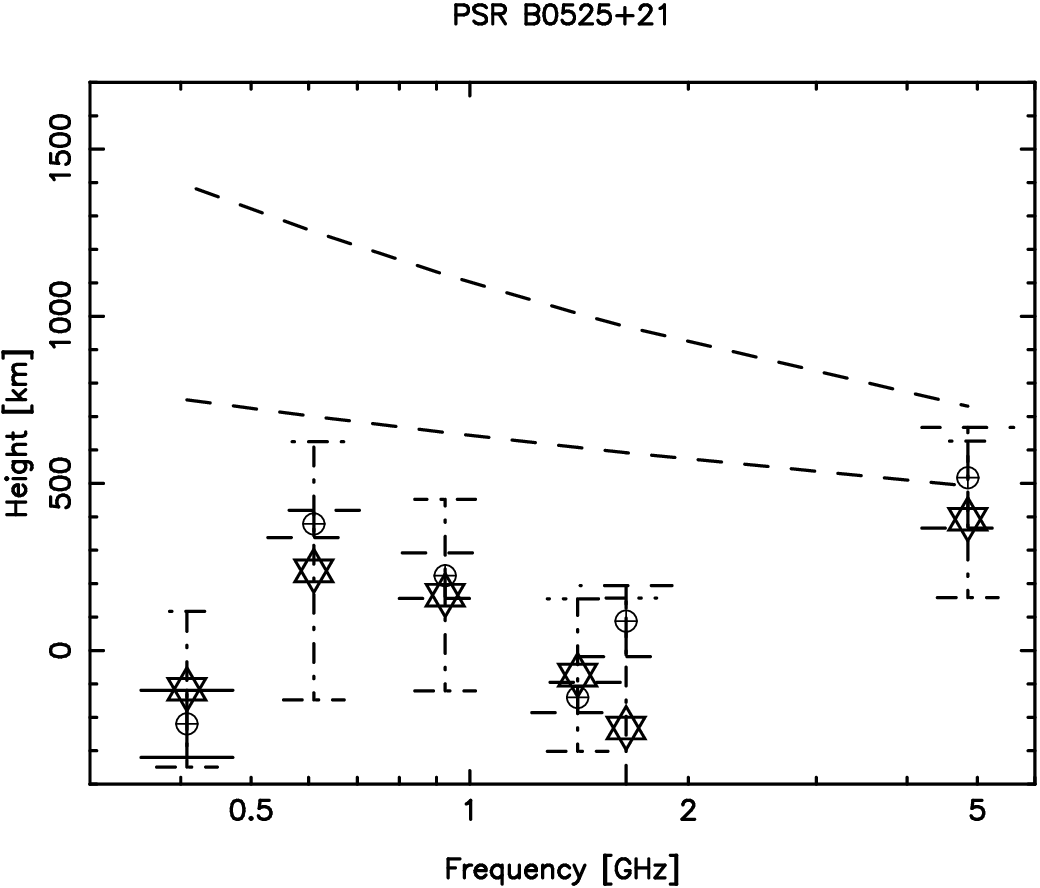}}}&
{\mbox{\includegraphics[width=5cm,height=5cm,angle=-90]{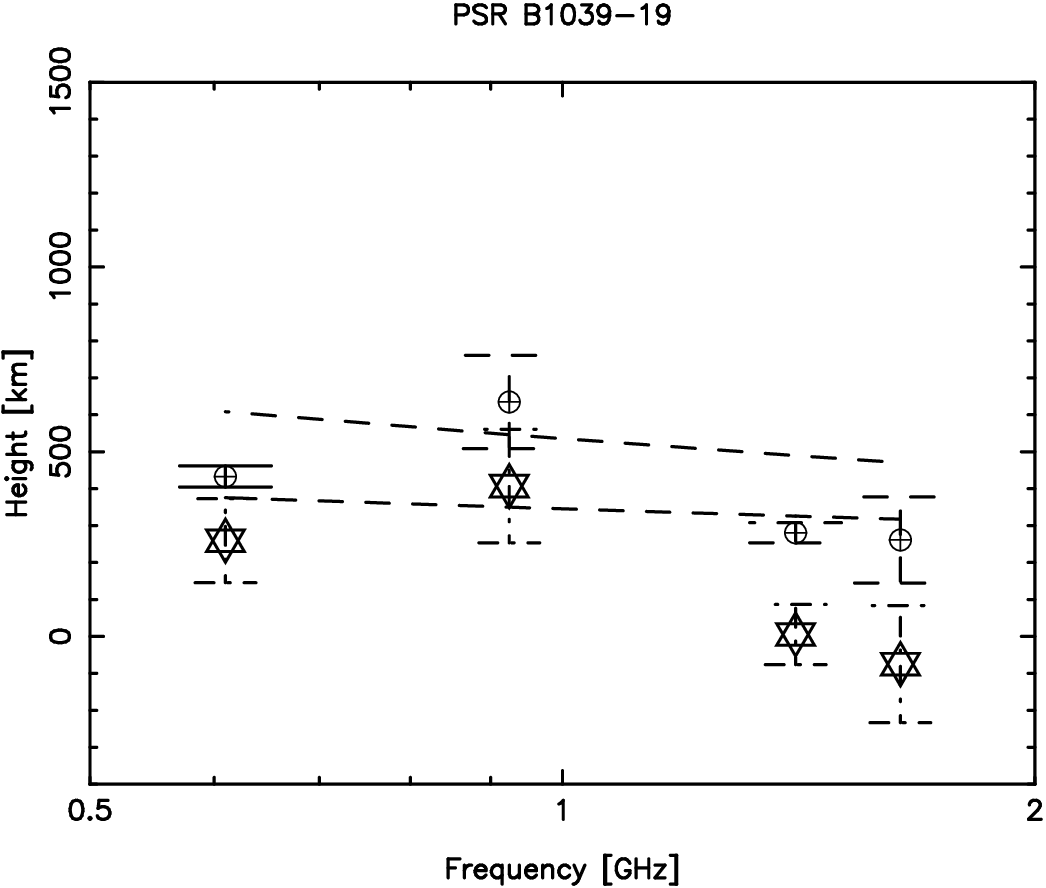}}}\\
{\mbox{\includegraphics[width=5cm,height=5cm,angle=-90]{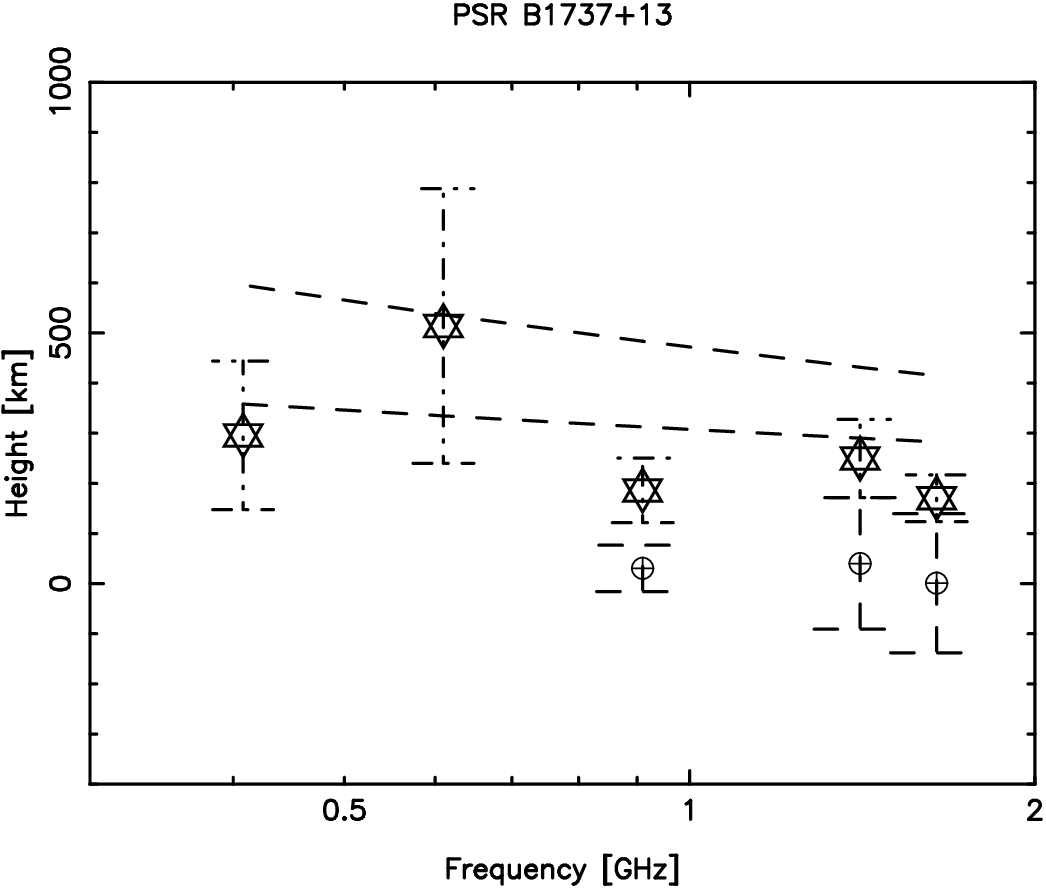}}}&
{\mbox{\includegraphics[width=5cm,height=5cm,angle=-90]{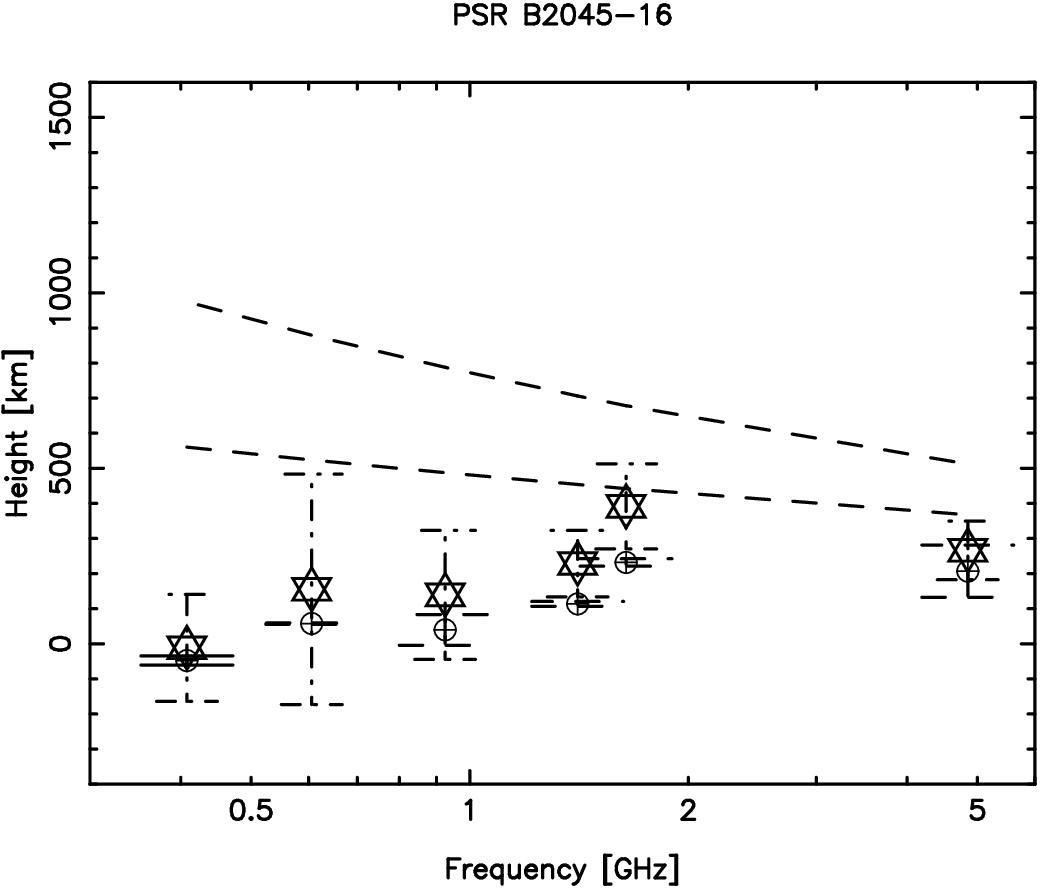}}}&
{\mbox{\includegraphics[width=5cm,height=5cm,angle=-90]{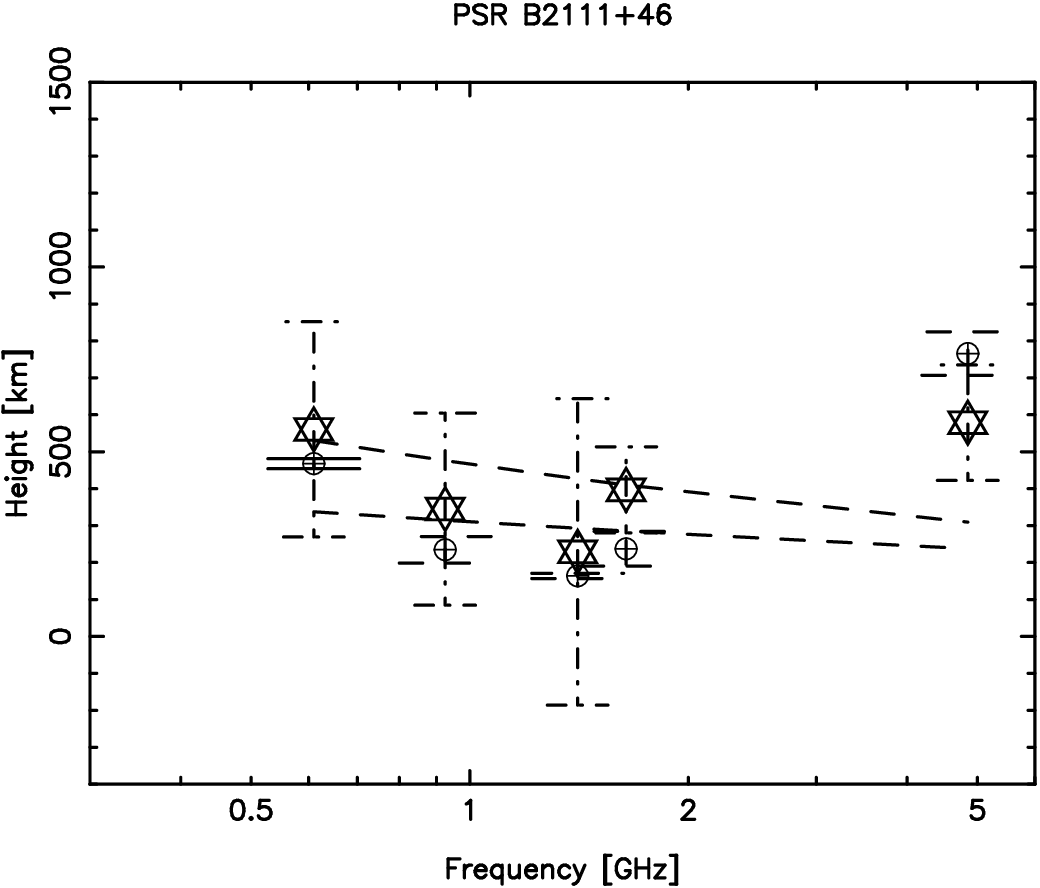}}}\\
\end{tabular}
\caption{In the figures the frequency dependence of $\rm r_{\rm delay}^{\rm BCW}$
for PPCS (circles with crosses) and OTS (stars)
measurements are shown.  The dashed line correspond
to upper and lower limits of the $\rm r_{geo}^{\rm KG}$ emission heights See text
for further detail.}
\label{fig3}
\end{figure*}

{\em The BCW method:} This method involves measuring $\Delta \phi =
\phi_{\rm profile} - \phi_{0}$ where $\phi_{\rm profile}$ is the
center of the pulse profile. Eventually the emission height is
obtained using the formula,
\begin{equation}
\rm{r_{delay}^{BCW}}\simeq-\frac{c}{4}\cdot\frac{\Delta\phi}{360^{\circ}}\cdot P~km.
\label{eq4}
\end{equation}
Here $\rm c=3\times 10^5$ km/s is the velocity of light.  The center
of the pulse profile $\phi_{\rm profile}$ requires accurate
measurement of the phase $\phi_{t}$ and $\phi_{l}$ at the extreme
outer trailing and leading edge of the profile i.e, $\phi_{\rm
profile} = \phi_{t}+(\phi_{l} - \phi_{t})/2$. The phase $\phi_{0}$
is the fiducial phase which is considered to be the steepest gradient point of the
PPA traverse. Note that $\rm r_{\rm delay}^{\rm BCW}$ corresponds
to the actual emission height if it is assumed that the emission
height corresponding to trailing, leading edge ($\rm r_{\rm edge}$)
and the inner ($\rm r_{\rm in}$) fiducial phase arises from the same
height.  For estimating $\rm r_{\rm delay}^{\rm BCW}$ one needs to use
phase points corresponding to lowest possible emission. In
figure~\ref{fig3} we show the comparison of $\rm r_{\rm delay}^{BCW}$
measured using PPCS and OTS measurements for all our sample of pulsars as
a function of frequency. We notice the following from the
figure,

(a) The frequency dependence of $\rm r_{\rm delay}^{\rm BCW}$ with
frequency is extremely erratic in contrast to the behaviour of $\rm
r_{\rm geo}$.  This erratic behaviour is present for both PPCS and OTS
methods.  Note that in several cases $\rm r_{\rm delay}^{\rm BCW}$ has
a negative value.

(b) $\Delta \phi$ measured using the PPCS and OTS methods gives values
of $\rm r_{\rm delay}^{\rm BCW}$ which agree well within the error
bars.

(c) $\rm r_{\rm delay}^{\rm BCW}$ measured using OTS method is
compatable with $\rm r_{\rm delay}^{\rm BCW} < \rm r_{\rm geo}^{\rm
KG}$ or $\rm r_{\rm geo}^{\rm OTS}$ as seen from table~\ref{tab2} and
\ref{tab3} (except for the 4.85 GHz point of PSR B2111+46).

To explain the erratic behaviour (point (a) above) of $\rm r_{\rm
delay}^{\rm BCW}$ with frequency one first needs to critically assess
factors affecting determination of $\Delta \phi$ i.e. the phase points
$\phi_{t}$, $\phi_{l}$ and $\phi_{0}$.
If the intensity distribution
of the outer edge of the pulsar beam is irregular (or jittery), this will
affect determination of $\phi_{\rm profile}$. The reason is 
$\phi_{t}$ and $\phi_{l}$ points at various frequencies will no longer
be symmetric with respect to the fiducial phase
$\phi_{0}$ which is breakdown of the key assumption of the BCW method. 
Conclusions drawn from the earlier sections suggest that
PPCS, OTS and OTHX methods within the error bars agree with a smoothly
varying profile edge. High quality observations are needed to discern
this effect. Incorrect choice of the fiducial phase $\phi_{0}$ can 
also result in the observed erratic behaviour. However as mentioned in
section~\ref{sec3} we have beyond doubt identified the  steepest gradient point 
as the fiducial phase for our sample of pulsars.

The uncertainties in the phase point measurements are in principle 
random and thus a variety in the relation between $\rm r_{\rm
delay}^{\rm BCW}$ and $\rm r_{\rm geo}$ should be observed. On the
contrary the general tendency $\rm r_{\rm delay}^{\rm BCW} < \rm
r_{\rm geo}^{\rm KG}$ (or $\rm r_{\rm geo}^{\rm OTS}$) is seen. Thus
the obvious question is what other factors affect $\rm r_{\rm
delay}^{BCW}$?  Here we discuss a few possibilities:

-- Broad emission region: BCW in their pioneering work points out that
their method gives $\rm r_{\rm delay}^{BCW}$ which are averaged over
all emission regions.  Recently Dyks et al. (2003) provides a lucid
description of this effect. They noted that $\rm r_{\rm delay}^{\rm
  BCW}$ can be decomposed in two parts where due to rotation the
profile center shifts towards earlier phases by -2 $\rm r_{\rm
  edge}/\rm R_{\rm lc}$ and the center of the PPA traverse delays by
2$\rm r_{\rm in}/ {\rm R_{\rm lc}}$ (where $\rm R_{\rm lc}$ is the
radius of the light cylinder). For $\rm r_{\rm edge} = \rm r_{\rm
  in}$, the estimated emission height corresponds to the emission
height of the outer conal edge.  In case $\rm r_{\rm edge} >> \rm
r_{\rm in}$ the height obtained is less than the outer edge emission,
but at the most can differ by a factor of 2. We check this effect
first by comparing $\rm r_{\rm delay}^{\rm BCW}$ obtained by the OTS
method $\rm r_{\rm geo }^{\rm OTS}$.  The value $\rm r_{\rm geo}^{\rm
  OTS}/\rm r_{\rm delay}^{\rm OTS}$ (listed in coloumn 8 of
table~\ref{tab3}) shows a large spread and for several cases are seen
to be greater than 2.  Given this variety no simple conclusion can be
drawn regarding the relation between the inner and outer edge
emission.

The difficulty in the above comparison is that the $s$ factor
corresponding to the OTS points is unknown.  If we assume $\rm r_{\rm
edge} >> r_{\rm in}$, then we can equate $\rm r_{\rm geo}^{OTS}
s^{-2}=2r_{{\rm delay}^{\rm OTS}}$ and find the $s$ parameter. In
table~\ref{tab3} the $s$ parameter is given and a wide
variety in $s$ values are seen thus making it difficult to reach any
meaningful conclusion.  However $\rm r_{\rm
delay}^{\rm OTS}$ {\em can} be affected by several other effects of unknown
magnitude which makes it impossible to extract the $s$ parameter.

-- Magnetic field sweepback (MFSB): Magnetospheric rotation and
currents can give rise to magnetic field distortions at sufficiently
high emission altitude $\rm r$.  The effect acts in the direction
opposite to the stellar rotation.  Unfortunately the magnitude of this
effect is not cleary understood. Shitov (1983) estimates the effect to
be of the order of $(\rm r / \rm R_{lc})^3$ while Arendt \& Eilek (1998)
finds it to be proportional to $(\rm r / R_{\rm lc})$.  
Using more sophisticated calculations Dyks et al. (2003) shows that this
effect is as important as A/R i.e: of the order of $ x(\rm r/R_{\rm
  lc})$, where $x$ is close to unity, confirming results obtained 
by Arendt \& Eilek (1998). Their calculations
demonstrate that this effect can cause distortions in the outer edge
of the beam which further depends on the viewing geometry.

 Phase shifts due to MFSB and aberration effects act in opposite
directions.  As a result, increase in $\Delta \phi$ due to A/R is
cancelled by decrease in the phase shift due to MFSB. This
underestimates $\Delta \phi$ and hence $\rm r_{\rm delay}$ and thus
can explain why $\rm r_{\rm delay} < \rm r_{\rm geo}^{\rm
OTS}$. Further the numerical calculations of Dyks et al. (2003) shows
that the magnitude of the MFSB effect varies for outer edge of the
radio beam as a function of $\beta$ or different magnetic field line
corresponding to the outer edge. Thus for multifrequency observation
since the outer edges corresponds to different field lines, MFSB
effects can be frequency dependent. 

\begin{figure*}
\begin{tabular}{lr@{}lr@{}lr@{}}
{\mbox{\includegraphics[width=5cm,height=5cm,angle=-90]{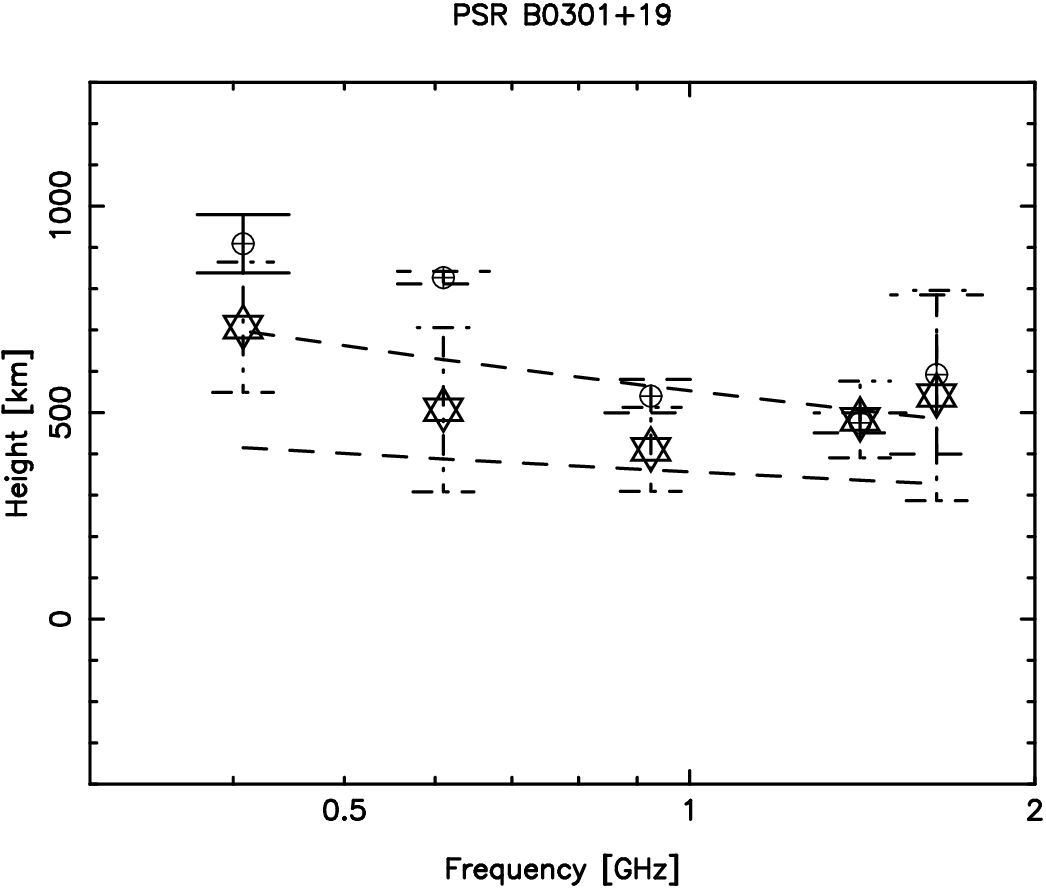}}}&
{\mbox{\includegraphics[width=5cm,height=5cm,angle=-90]{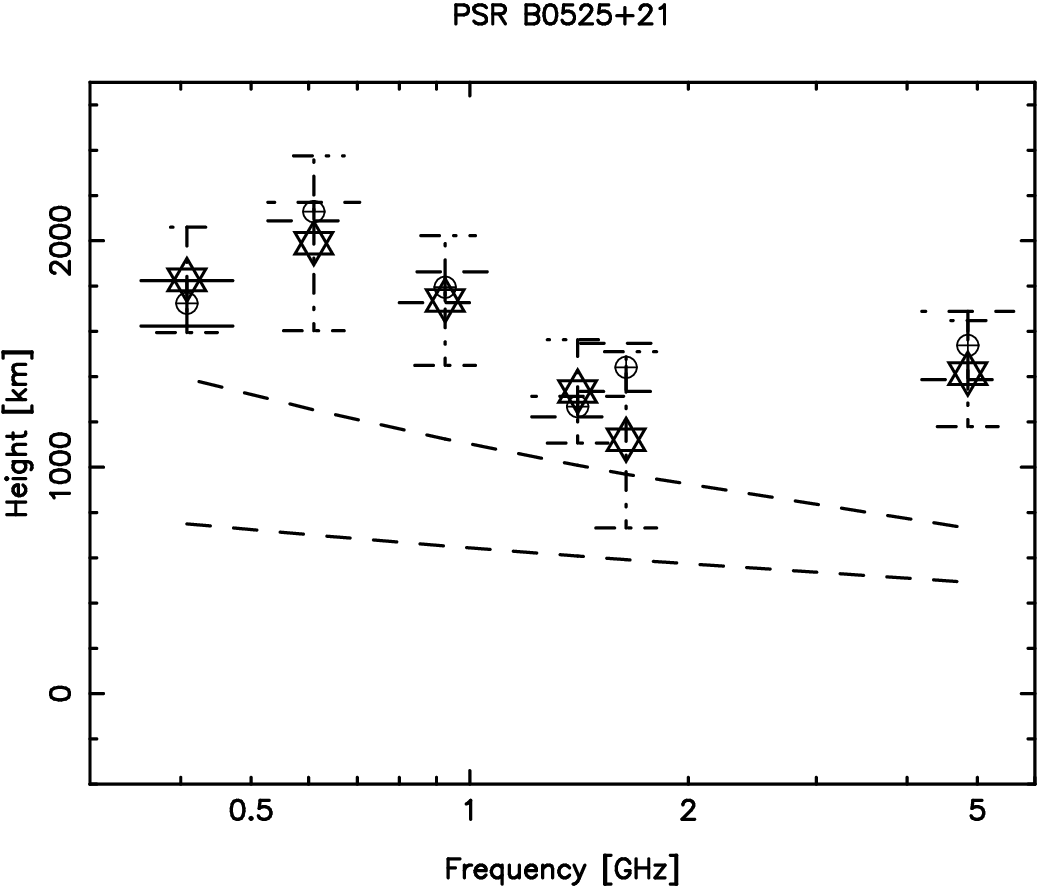}}}&
{\mbox{\includegraphics[width=5cm,height=5cm,angle=-90]{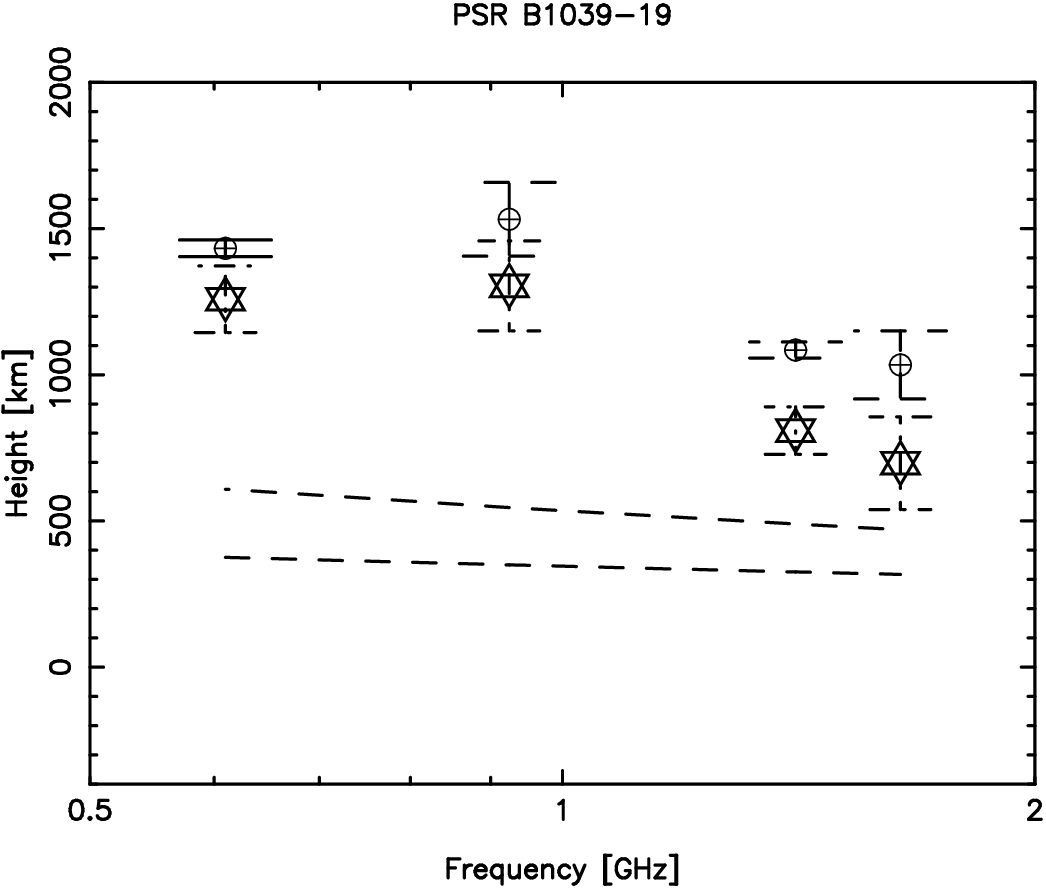}}}\\
{\mbox{\includegraphics[width=5cm,height=5cm,angle=-90]{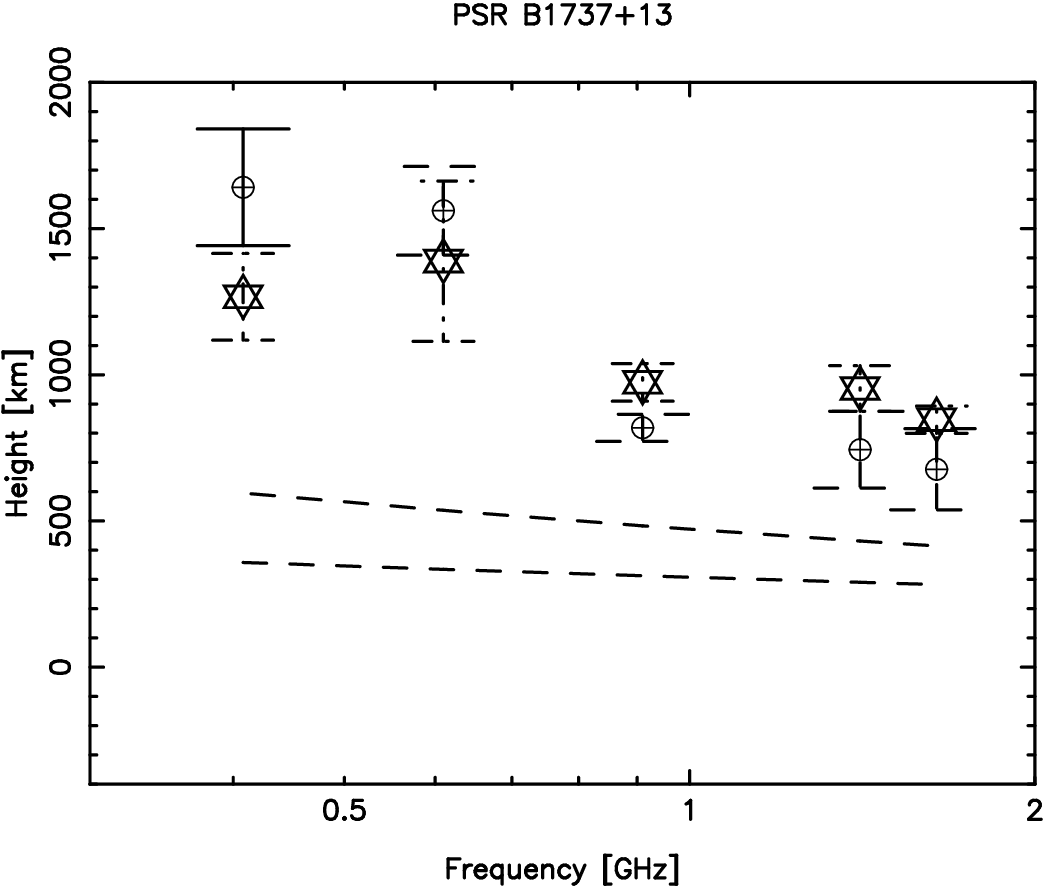}}}&
{\mbox{\includegraphics[width=5cm,height=5cm,angle=-90]{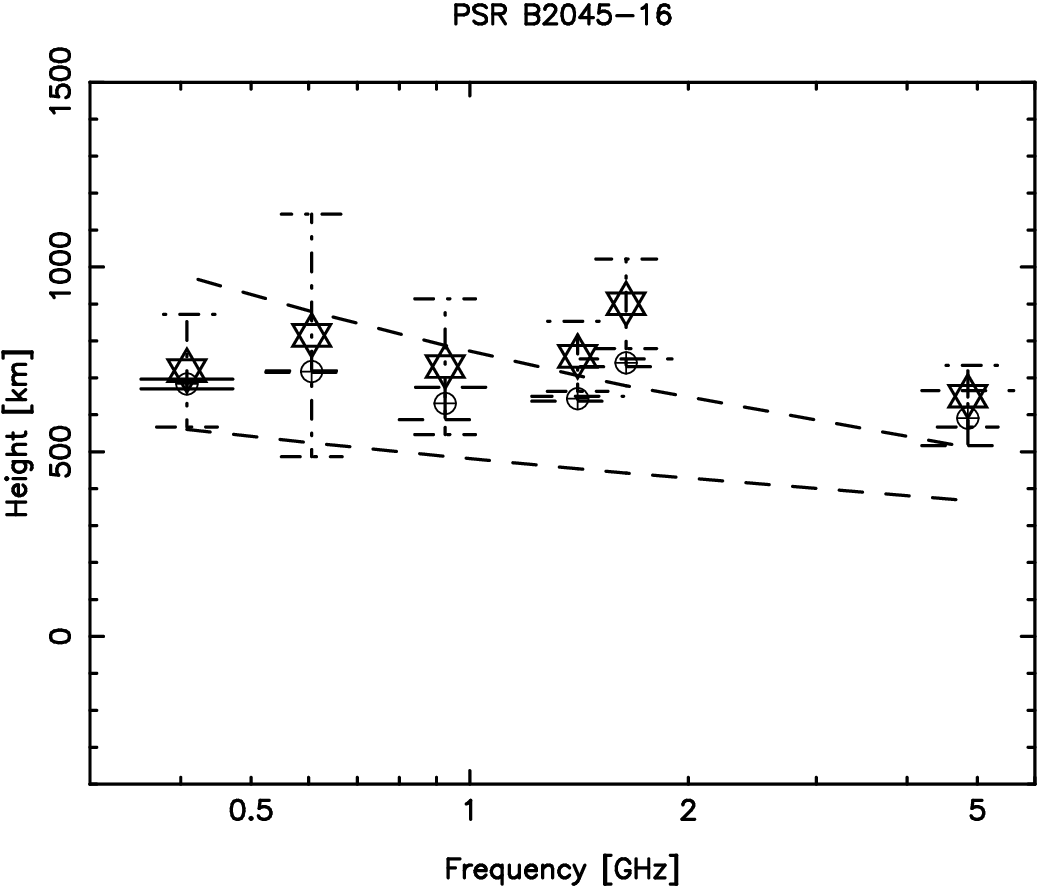}}}&
{\mbox{\includegraphics[width=5cm,height=5cm,angle=-90]{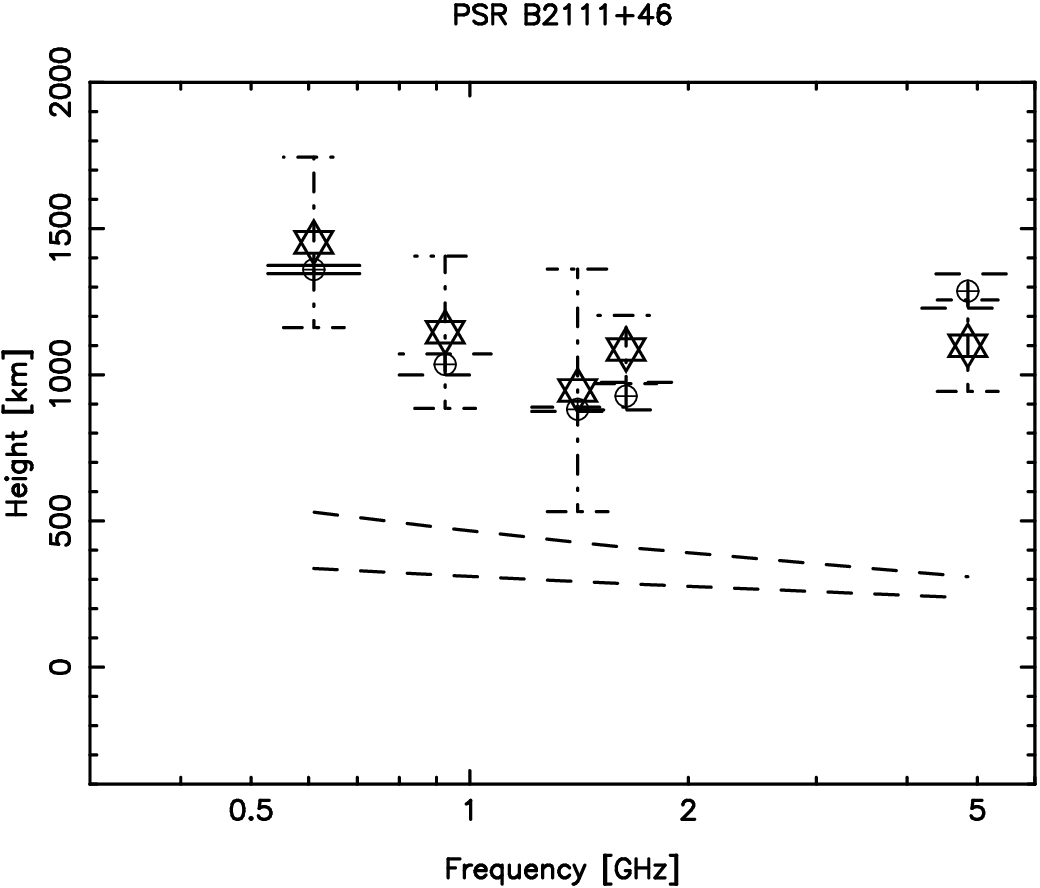}}}\\
\end{tabular}
\caption{In the figures the frequency dependence of $\rm r_{\rm delay}^{\rm BCW}$
corrected for a simulated MFSB effect for PPCS (circles with crosses) and OTS (stars)
measurements are shown.  The dashed line correspond
to upper and lower limits of the $\rm r_{geo}^{\rm KG}$ emission heights See text
for further detail.}
\label{fig4}
\end{figure*}

If we assume that all the emission arises from the same
height, and only A/R and MFSB effect contributes to $\Delta \phi$, we
can decompose $\Delta \phi = \Delta \phi_{\rm A/R} - \Delta \phi_{\rm
MFSB}$. Further if we claim that $\rm r_{\rm delay} = \rm r_{\rm
geo}^{\rm KG}$, then $\Delta \phi_{\rm MFSB}$
should be added to the observed $\Delta \phi$ in order to satisfy the
equality. This requires $\Delta \phi_{\rm MFSB} =  x (4 \times
360^{\circ}\times {\rm c}\times {\rm r}_{\rm geo}^{\rm KG}/ {\rm P})$, where the factor $x$ is
not known.  For $x\sim1$ one can in principle get fairly good
agreement with the delay and the geometrical heights as seen in figure~\ref{fig4}. 
If the MFSB effect is indeed operating in pulsars then we have found 
observational confirmation of this for the first time.

Given the multitude of effects of unknown magnitude
simultaneously affecting $\rm r_{\rm delay}^{\rm BCW}$ the height
estimates obtained from this method are affected by a factor of few.

\begin{figure*}
\begin{tabular}{lr@{}lr@{}}
{\mbox{\includegraphics[width=5cm,height=8cm,angle=-90]{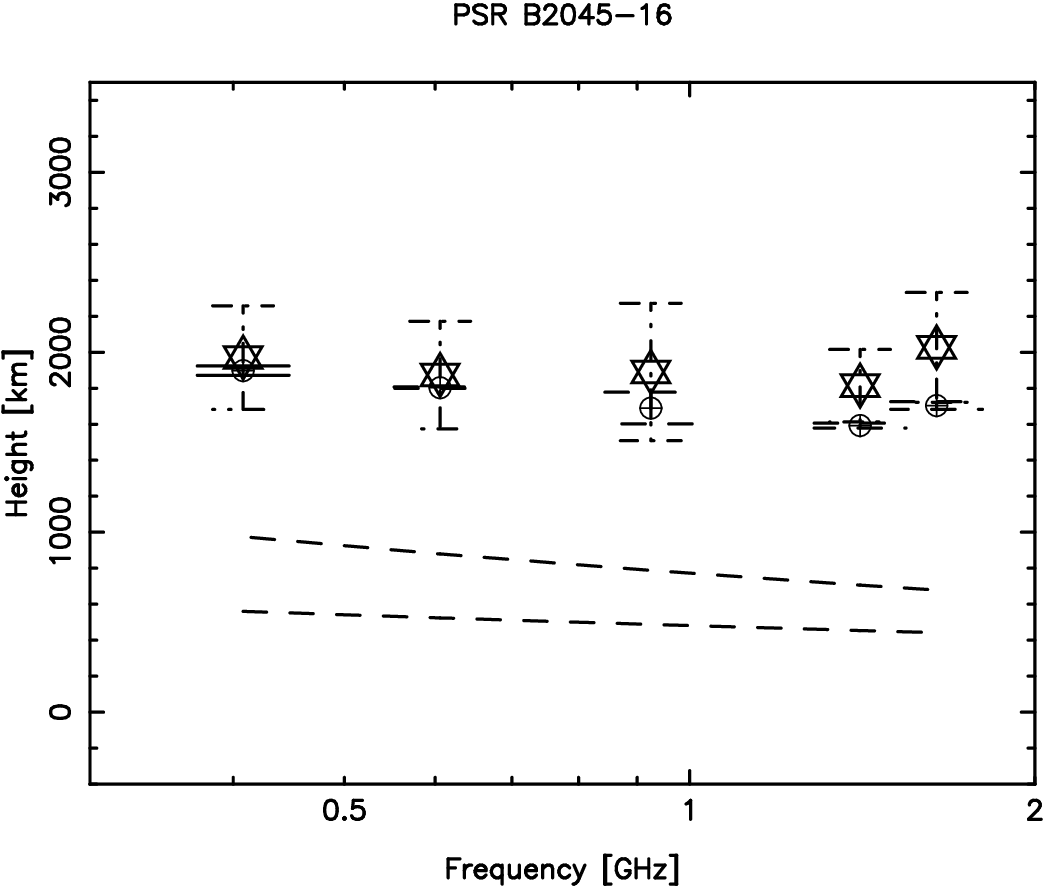}}}&
{\mbox{\includegraphics[width=5cm,height=8cm,angle=-90]{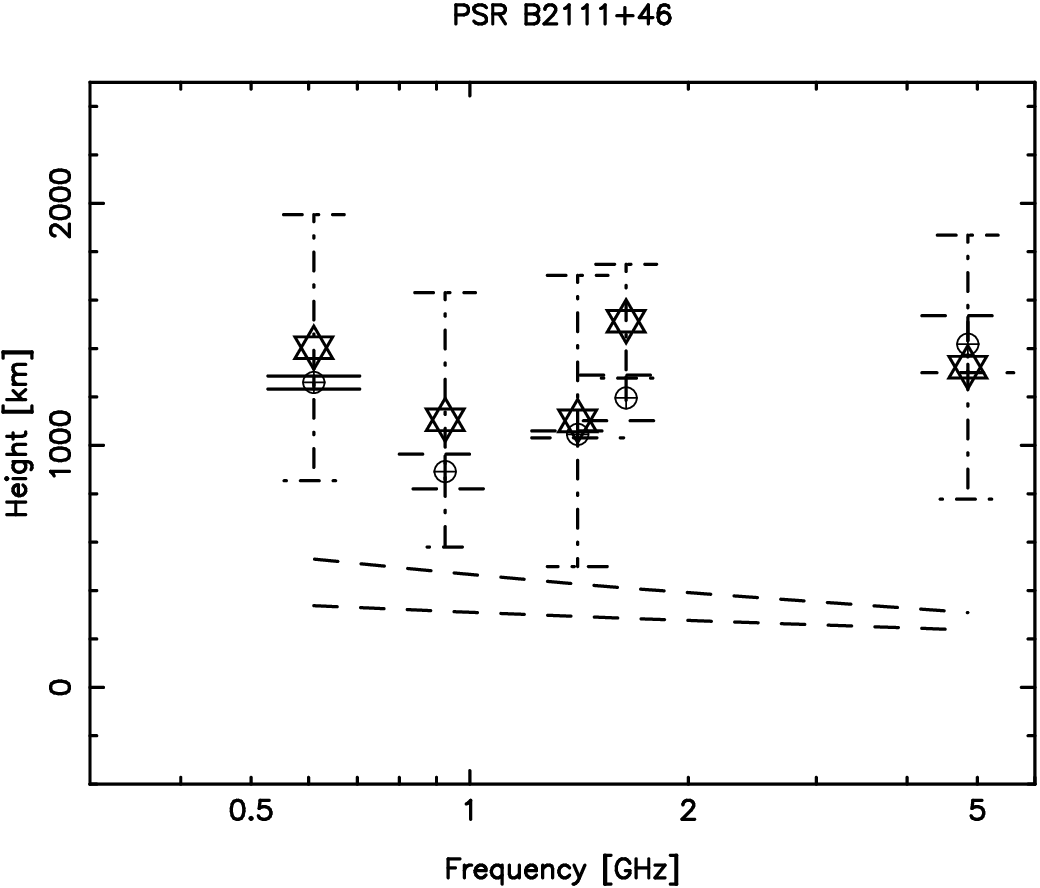}}}\\
\end{tabular}
\caption{In the figures the frequency dependence of $\rm r_{\rm delay}^{\rm GGa}$
for PPCS (circles with crosses) and OTS (stars)
measurements are shown.  The dashed line correspond
to upper and lower limits of the $\rm r_{geo}^{\rm KG}$ emission heights See text
for further detail.}
\label{fig5}
\end{figure*}

{\em The GGa method:} This method is also based on the A/R effect
suggesting an alternative way of deriving delay emission heights based
on total intensity profiles.  The method applies to pulsars with both
core and conal emission. Here the fiducial phase is assumed to be
the peak of the core emission. 
The emission heights obtained
are heights with respect to the height of the core emission ($\rm r_{\rm core}$).  

GGa's method involves measuring the phase shift $\Delta\phi$
between the center obtained from the PPCS and the peak of the core
emission. The emission height corresponding to the peak of the core
emission ($\phi_{\rm c}$) and the peaks of the conal components
$\phi_{\rm l}$ and $\phi_{\rm t}$.  GGa connects $\Delta\phi$ to the
emission radius as $\rm r_{\rm delay}^{\rm GGa} = (\rm c/2) (\rm P / 360^{\circ}) \Delta\phi / (1 +
\sin \alpha)$. An important correction to this formula has been
suggested by Dyks et al (2003) where they find,
\begin{equation}
\rm r_{\rm delay}^{GGa} = - \frac{c}{2} \frac{\Delta\phi}{360^{\circ}} P~km. 
\label{eq5}
\end{equation}
Note that the modified formula is independent of the emission
geometry.  GGa and GGb applied their method to seven M
type pulsars at 325 MHz. They demonstrated that the centers of both
outer and the inner emission components lead the peak of the core
emission supporting the A/R picture. They also found $\Delta\phi$ for
the inner components are systematically smaller than the outer
components, suggesting that the inner components arise closer to the
stellar surface compared to the outer components and the core emission
lies below the conal emission and closest to the stellar
surface. Their observation supports the picture that the pulsar
emission beam has nested cones of emission (e.g. Mitra \& Deshpande
1999) with the inner cones arising at lower heights compared to the
outer cones as was suggested by Rankin (1983).

Considering the above scenario if $\rm r_{\rm core}= R_{*}$, a
straightforward comparison can be made between $\rm r_{\rm delay}^{\rm
GGa}$ and $\rm r_{\rm geo}$.  If $\rm r_{\rm delay}^{\rm GGa} \sim
r_{\rm geo}$, then one can conclude that the field line corresponding
to OTS points are the last open field lines such that $s=1$. In the
case $\rm r_{\rm delay}^{\rm GGa} > \rm r_{\rm geo}$ the OTS points
correspond to $s<1$, giving lower values of $\rm r_{\rm geo}^{\rm
OTS}$. Thus claiming that $\rm r_{\rm delay}^{\rm GGa}$ should be
equal to $\rm r_{\rm geo}^{\rm OTS}$ and using equation~\ref{eq2} one can
find $s=(r_{geo}^{\rm OTS}/r_{\rm delay}^{GGa})^{0.5}$.  We use PSR
B2045-16 and PSR B2111+46 in our data set where such a comparison is
possible.  In table~\ref{tab4} values for $\rm r_{\rm delay}^{\rm
GGa}$ and the $s$ parameter for these two pulsars are given. In
figure~\ref{fig5} $\rm r_{\rm delay}^{\rm GGa}$ is shown.  For both the
cases one observes

(a) $\rm r_{\rm delay}^{\rm GGa} > \rm r_{\rm geo}^{\rm OTS}$ and $s <
1$ as given in table~\ref{tab4}.

(b) $\rm r_{\rm delay}^{\rm GGa}$ obtained by PPCS and OTS methods
agrees within the error bars.

(c) Variation of $\rm r_{\rm delay}^{GGa}$ with frequency is erratic.
For PSR B2045-16 the variation is different from that obtained by the
BCW method.

(d) For both the cases $\rm r_{\rm delay}^{\rm GGa} > \rm r_{\rm
delay}^{\rm BCW}$.

A crucial assumption in the GGa method is considering the peak
of the core emission as the fiducial phase containing the 
dipole axis, the rotation axis and the observers line of sight.
GGa already mentions that if $\rm r_{\rm core}$ originates higher than the polar cap,
then the height of the core needs to be added to $\rm r_{\rm delay}^{\rm GGa}$
to obtain correct radio emission heigths. Nonetheless even in this case
the peak of the core emission needs to be the fiducial phase as defined
above.

\begin{figure}
\begin{center}
\begin{tabular}{@{}lr@{}}
{\mbox{\includegraphics[width=8cm,height=8cm,angle=-90]{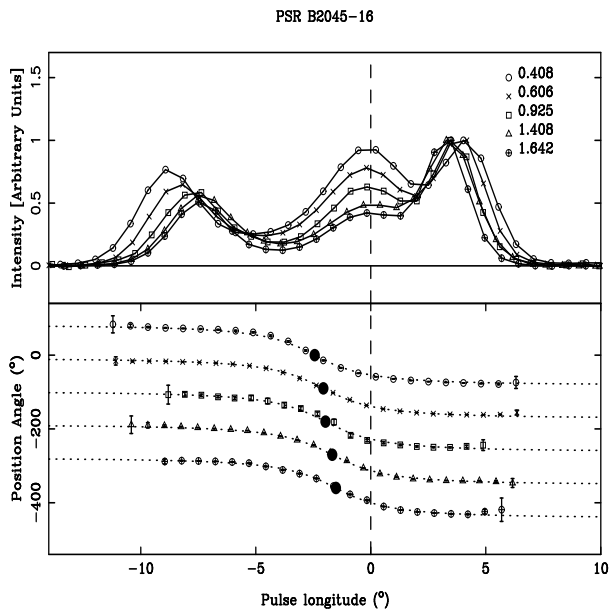}}}
\end{tabular}
\caption{The above plots shows the multifrequency alignment of 
PSR B2045$-$16 based on the peak
of the core component. The top panel shows the intensity and the 
bottom panel the PPA traverse. The dark dots in the 
bottom panel corresponds to the steepest gradient point in 
the PPA traverse (see text for details).} 
\label{fig6}
\end{center}
\end{figure}

In section~\ref{sec3} based on our combined PPA traverses we have 
firmly concluded that the steepest gradient point in the PPA traverse
serves as the fiducial phase in our sample of pulsars. Thus 
for the core to originate from the polar cap and 
the peak of the core to be the fiducial phase it should be coincident with the 
steepest gradient point. Here we notice that
for PSR B2045$-$16 and PSR B2111+46 even by eye inspection (see
figure~\ref{fig1} as examples) it is clearly seen that the center of
the core lags the steepest gradient of the combined PPA traverse. This feature
is also evident in PSR B1737+13 at frequencies 408 and 610 MHz where
the core component is clearly identifiable.  In figure\ref{fig6} we
show in the top panel the alignment of PSR B2045-16 with respect to
the core emission. The core emission clearly lags the steepest
gradient point of the PPA traverse shown in the lower panel and the
amount of lag is seen to increase with decreasing frequency.

The central core emission is long thought to originate from the
surface of the polar cap (Rankin 1990, 1993).  Rankin noticed that the
3 dB core widths at 1 GHz for six interpulsars depend on the pulsar
period and $\alpha$ and this relates extremely well with the size of
the polar cap estimated using dipolar field lines. It is hence
conjectured that the core emission is consistent with emission arising
from a filled polar cap with a bivarient Gaussian intensity
distribution, and the peak of the core lies in the fiducial plane
containing the dipolar magnetic axis.  If the core emission arises
from the polar cap then: (1)
According to the BCW model the phase offset between the core peak and
the steepest gradient of the PPA traverse is $\rm (4/c) R_{*}
(360^{\circ}/P) $. For a typical 1 sec pulsar and $\rm R_{*}=10$ km this
number is extremely small $\sim 0.05^{\circ}$, typically a factor of
10 less than the typical resolution of observations. On the other hand
the observed lag of the core emission with respect to the steepest
gradient point is in disagreement with the core emission arising from
the polar cap.  (2) Further if the core emission arises from the polar
cap at all frequencies the width of the core should be frequency
independent.  However as seen in figure~\ref{fig6} the core width
evolves significantly with frequency. (3) The line-of-sight $\beta$
should be smaller than the polar cap radius $\rm \rho_{pc} (^{\circ})=
1.24 R_* P^{-0.5}/10$.  In table~\ref{tab1} values of $\rho_{\rm pc}$ is
given (for $\rm R_* = 10\pm3$ km) and for the core pulsars $\rho_{\rm pc} \sim \beta$ indicating
that the line of sight is almost grazing the polar cap rather than
central cuts.  Also Dyks et al. (2003) finds $\rho_{\rm pc} > \beta$
for the core dominated pulsar PSR B0450$-$18. These evidences hint that the
core emission does not originate from the polar cap surface and the
peak of the core is not the fiducial phase. Breakdown of this assumption 
introduces unknown uncertainties in $\rm r_{\rm delay}^{\rm GGa}$ 
and thus comparison with other emission height estimates are difficult.
\footnote{Note that we have based our arguments accepting that by combining
the PPA traverse and establishing its striking agreement with the RVM, 
we have found the fiducial phase as the steepest gradient point.   
Any other alternative explanation refuting the steepest gradient
point to be the fiducial phase will affect the conclusion drawn
in this paper.}

It is however puzzling that if the peak of the core is not the fiducial 
phase why all the seven pulsar subject to study by GGa and GGb has the 
core lagging the profile center.  Lyne \& Manchester
(1988) already noted that they find 58 core dominated pulsars
for which `35 lead the profile center and 25 trail it'. 
Perhaps this observational feature needs to be reviewed before we can 
reach a firm conclusion.
It should be remembered the MFSB effect as discussed earlier also affects
determination of $\rm r_{\rm delay}^{\rm GGa}$ in a very similar fashion.

\subsection{Effect of propagation on emission heights:}

Propagation effects within the pulsar magnetosphere can influence
polarization and total intensity properties in pulsars as was
advocated by Barnard \& Arons (1986) and recently reinvestigated in a
series of work by Petrova \& Lyubarskii (2000), Petrova (2000) and
Petrova (2003 and references therein).  The effect of propagation can
complicate estimation of $\rm r_{\rm geo}$.  This arises in the form
of refractive effects acting on the outer conal components.
Refraction causes the rays to decline the outer conal component more
outward.  The effect is frequency dependent and is prominent at higher
frequencies and is considered as an explanation for saturation of
pulse widths (see MR02) observed at high frequencies (Mckinnon 1997).
However the magnitude of this effect depends on the plasma parameters
in the pulsar magnetosphere which is difficult to constrain.  However
as a consequence the width $\rm W$ is larger and $\rm r_{\rm geo}$
corresponds to emission heights which are higher than the actual
emission heights.

The effect of propagation can also affect determination of $\rm r_{\rm delay}$. 
Unfortunately the
theory for such effects are still not well understood.  Grossly it is
conjectured that in the course of propagation through the plasma flow
the PPA follows the direction of the local magnetic field upto a
distance $\rm r_{\rm p}$, higher than the total intensity emission.
According to the BCW model the effect will tend to increase $\Delta \phi$,
since the the steepest gradient will be delayed by a larger amount
compared to the total intensity profile. This should however result in
higher estimates of $\rm r_{\rm delay}$ compared to $\rm r_{\rm geo}$,
contrary to what is observed.  Stongly varying $\rm r_{\rm p}$ across
the profile can complicate interpretation of $\rm r_{\rm delay}$
significantly. This can also affect phase determination of the beam
edge in a frequency dependent manner explaining the erratic nature of
$\rm r_{\rm delay}$.

Further PPA's are influenced mostly in regions in the pulse where the
amount of circular polarization is high. A general feature observed in
average profiles is that cores are more circularly polarized than the
cones, and thus PPA's will tend to be more affected near the core
component. Single pulse analysis have recently shown that circular
polarization is a property of both cone and core pulsars (Karastergiou
et.  al. 2003). The averaging effect washes out the circular
polarization in the cones, however retains them for the central core
emission. Thus in average profiles for outer cones propagation effects
might be averaged out, but can still influence the core
component. Although this can influence determination of $\phi_{0}$ and
consequently $\rm r_{\rm delay}$. Nonetheless the remarkable similarity
of the PPA traverses over frequency seems to rule out any major
propagation effect that might influence PPA traverses 
in our sample of pulsars.

\section{Summary}
\label{sec5}

Important conclusions drawn in this paper are the following:

(a) For the first time using six bright conal pulsars
we construct the combined PPA
traverse ranging from frequencies between $\sim$ 400 MHz to $\sim$ 5
GHz. The combined PPA traverse is in excellent agreement with the RVM 
confirming that radio emission arises from regions with perfectly dipolar 
magnetic field.

(b) A fiducial phase in a pulse is chosen
to be the point lying in the plane containing the dipolar magnetic axis, 
the rotaion axis and the observers line of sight. 
Due to the retardation effect (BCW) and RFM this fiducial plane 
changes with frequency.
By combining the PPA traverses we eliminate this frequency dependent effect
and constructed a common fiducial plane containing the fiducial phase
(see dark dots in the bottom panel of figure~\ref{fig6} or/and dashed 
vertical lines in figure~\ref{fig1}).
Then noting the striking similarity with 
the RVM model we conclude that the steepest gradient point of the combined
PPA traverse in the fiducial phase for our sample of pulsars. This phase
thus should be treated as the fiducial phase for the application
of cold plasma dedispersion and timing analysis.

(c) We find in general $\rm r_{delay}^{\rm BCW} < \rm r_{geo}^{\rm OTS}$
and the frequency dependence of $\rm r_{delay}^{\rm BCW}$ is more erratic
than $\rm r_{geo}^{\rm OTS}$. The former effect can be understood by
considering a broad emission region (this means emission at any 
given frequency arises over a range of heights in the pulsar 
magnetosphere) and/or the MFSB effect recently proposed
by Dyks et al. (2003). The latter effect can arise due to irregular (or jittery)
shape of the edge of the pulsar beam.

(d) The peak of the core emission for three pulsars is 
seen to lag the steepest gradient point
of the PPA traverse over several frequencies. Further the width of the core
is seen to increase with decreasing frequency. This strongly suggests
that the core emission emanates higher than the polar cap surface and the 
peak emission does not lie in the fiducial plane containing the dipole axis, rotation axis
and the line of sight.

In order to understand issues relating to pulse emission heights
future studies producing combined PPA traverses using lower frequency
observations ($<$400 MHz) should enable us to get better constraints on the viewing
geometry. On the theoretical side there is a great need to develop and understand the 
details of the MFSB effect. Futher efforts should be made to understand
the observed lag between the steepest gradient point and the peak of the 
core emission which has been highlighted in great detail in this work.

\acknowledgements X.H. Li thanks Prof. G.J. Qiao and X.H. Sun for
valuable discussion.  We thank J. Gil, S. Petrova,
J. H. Seiradakis, M. Kramer, D. Athanasiadis, Y. Gupta and J. Kijak
for insightful discussions and comments on the manuscript. We thank
J. L. Han, A. Jessner \& R. Wielebinski for constant encouragement
during this work and several useful comments.  We thank J. Rankin for
critical comments on this work which helped us to improve the
manuscript substantially.  The project is supported by the National
Natural Science Foundation of China (19903003 and 10025313), the
National Key Basic Research Science Foundation of China (G1990754) and
the partner group of MPIFR at NAOC.

\newpage

\begin{table*}
\caption{The table below lists the pulsar at several frequencies subject to our current analysis. 
The angles $\alpha$, $\beta$ are fits obtained from the RVM model at every frequency and the
$\chi^{2}$ for the fits are given. The 
steepest gradient is designated by $(\frac{d\psi}{d\phi})_{max}$. The various widths ($\rm W$) 
and the geometrical radius $\rm r_{\rm geo}$ estimates for the OTS, OTHX and PPCS points are tabulated. \label{tab2}}
\begin{tabular}{c@{\hspace{0.15cm}}c@{\hspace{0.15cm}}r@{$\pm$}l@{\hspace{0.15cm}}r@{$\pm$}l@{\hspace{0.15cm}}c@{\hspace{0.15cm}}r@{$\pm$}l@{\hspace{0.15cm}}r@{$\pm$}l@{\hspace{0.15cm}}r@{$\pm$}l@{\hspace{0.15cm}}r@{$\pm$}l@{\hspace{0.15cm}}r@{$\pm$}l@{\hspace{0.15cm}}r@{$\pm$}l@{\hspace{0.15cm}}r@{$\pm$}l}
\noalign{\smallskip} 
\hline
\hline
\noalign{\smallskip} 
PSR B&Freq&\multicolumn{2}{c}{$\alpha$}&\multicolumn{2}{c}{$\beta$}&$\chi^{2}$ &\multicolumn{2}{c}{$(\frac{d\psi}{d\phi})_{max}$}&\multicolumn{2}{l}{${\rm W}_{\rm OTS}$}&\multicolumn{2}{l}{${\rm W}_{\rm OTHX}$}&\multicolumn{2}{l}{${\rm W}_{\rm PPCS}$}&\multicolumn{2}{c}{$\rm r_{\rm geo}^{\rm OTS}$}&\multicolumn{2}{c}{$\rm r_{\rm geo}^{\rm OTHX}$}&\multicolumn{2}{c}{$\rm r_{\rm geo}^{\rm PPCS}$}\\
     &(GHz)&\multicolumn{2}{c}{($^\circ$)}&\multicolumn{2}{c}{($^\circ$)}&   &\multicolumn{2}{c}{}
&\multicolumn{2}{c}{($^\circ$)}&\multicolumn{2}{c}{($^\circ$)}&\multicolumn{2}{c}{($^\circ$)}
&\multicolumn{2}{c}{(km)}&\multicolumn{2}{c}{(km)}&\multicolumn{2}{c}{(km)}\\\hline 		
\noalign{\smallskip} 			             						
B0301+19&  0.408&168&299&$-$0.6&14.0& 0.4&$-$18& 28&18.9& 0.5& 18.5& 0.5& 11.57& 0.04&  122 & 159&  118&  153& 53&   60\\
        &  0.610&166& 63&$-$0.7& 3.4& 3.2&$-$17&  9&22.9& 0.7& 22.0& 0.7&  9.76& 0.01&  174 & 233&  162&  215& 41&   44\\
        &  0.925&158&106&$-$1.0& 4.5& 0.1&$-$23& 24&19.2& 0.3& 18.7& 0.3&  8.73& 0.01&  126  & 164&  120& 156& 34&   36\\
        &  1.408&133& 49&$-$2.3& 1.9& 1.3&$-$18& 12&19.0& 0.3& 18.8& 0.3&  7.91& 0.01&  124  & 162&  121& 157& 31&   30\\
        & 1.642&158&2323&$-$1.0&118& 0.1&$-$23&408&15.1& 0.6& 14.8& 0.6&  7.62& 0.04&   82  & 101&  79&  98& 29&    28\\
	\multicolumn{21}{c}{}\\
B0525+21&  0.408& 14& 38&   0.3& 0.8& 1.8&   41& 15&21.4& 0.3& 21.4& 0.3& 14.61& 0.01& 1946& 616 & 1946&615& 928&  292\\
        &  0.610& 47&  7&   1.1& 0.2& 1.1&   37&  7&23.2& 0.5& 23.1& 0.5& 13.73& 0.01& 2269& 722 & 2247&715& 825&  259\\
        &  0.925& 81&  9&   1.7& 0.1& 0.6&   33&  2&21.2& 0.3& 21.1& 0.2& 13.19& 0.01& 1909& 605 & 1880&594& 764&  240\\
        &  1.408& 40& 10&   1.1& 0.3& 0.5&   33& 10&22.0& 0.3& 21.7& 0.3& 12.52& 0.01& 2049& 649 & 2005&634& 693&  217\\
        &  1.642& 73& 17&   1.6& 0.2& 0.5&   33&  5&21.4& 0.5& 20.7& 0.5& 12.11& 0.01& 1953& 622 & 1819&579& 651&  205\\
        &  4.850& 50& 21&   1.5& 0.5& 1.3&   29& 13&18.2& 0.2& 17.8& 0.1& 11.13& 0.01& 1413& 447 & 1316&431& 556&  175\\
	\multicolumn{21}{c}{}\\
B1039$-$19&0.610& 50& 32&$-$2.7& 1.2& 0.7&$-$16&  1&20.4& 0.4& 20.3& 0.4& 11.62& 0.02&  658&  379&  650&374& 252&   130\\
        &  0.925& 67&126&$-$3.6& 3.2& 0.1&$-$14& 23&18.1& 0.3& 17.9& 0.3& 10.75& 0.03&  529&  299&  518&293& 224&   113\\
        &  1.408&  7&278&$-$2.8& 0.1& 3.4&$-$21&  1&18.0& 0.3& 17.9& 0.2& 10.14& 0.02&  525&  297&  519&293& 206&   102\\
        &  1.642& 71&189&$-$3.0& 3.0& 0.6&$-$18& 27&17.5& 0.4& 17.4& 0.4&  9.72& 0.03&  501&  282&  494&278& 194&   95\\
	\multicolumn{21}{c}{}\\
B1737+13&  0.408& 84&109&$-$4.2& 0.8& 1.3&$-$14&  4&22.1& 0.6& 18.2& 0.8&\multicolumn{2}{c}{--}&   359&  175&  252&122&\multicolumn{2}{c}{--} \\
        &  0.610& 54& 20&$-$2.7& 0.7& 5.2&$-$17&  6&28.4& 1.5& 23.2& 0.4&\multicolumn{2}{c}{--}&   581&  293&  397&193&\multicolumn{2}{c}{--} \\
        &  0.910& 85& 58&$-$2.4& 0.2& 0.8&$-$24&  3&20.8& 0.3& 20.1& 0.3& 16.70&0.08&   322&  155&  302&145&216 &101 \\
        &  1.408& 38& 23&$-$1.6& 0.8& 2.4&$-$22& 16&24.6& 0.6& 23.4& 0.3& 15.88&0.08&   443&  217&  401&195&198 &84  \\
        &  1.642& 32& 58&$-$1.5& 2.3& 1.9&$-$21& 47&22.3& 0.3& 21.5& 0.3& 15.24&0.04&   367&  177&  345&166&184 &92  \\
	\multicolumn{21}{c}{}\\
B2045$-$16&0.408&111&  8&$-$1.5& 0.1& 0.5&$-$36&  3&19.4& 0.4& 19.2& 0.3& 12.88& 0.01&  810&  245&  792&239& 366&   110\\
        &  0.606&129&  2&$-$1.4& 0.1&31.1&$-$32&  1&22.7& 0.3& 22.3& 0.4& 12.09& 0.01&  1107& 334& 1066&322& 325&   97\\
        &  0.925&136& 53&$-$1.1& 1.1& 0.5&$-$36& 50&17.8& 0.4& 17.5& 0.4& 11.28& 0.02&  688&  209& 661& 200& 285&   86\\
        &  1.408&121&  6&$-$1.4& 0.1&10.2&$-$35&  3&19.6& 0.2& 19.3& 0.2& 10.84& 0.01&  829&  250&  805&242& 264&   80\\
        &  1.642& 99&  9&$-$1.7& 0.1& 4.3&$-$34&  2&18.8& 0.3& 18.5& 0.3& 10.71& 0.01&  763&  230&  737&222& 258&   78\\
        &  4.850& 92& 77&$-$2.4& 0.3& 2.5&$-$24&  4&14.7& 0.1& 14.3& 0.1& 10.11& 0.01&  471&  141&  445&134& 232&   70\\
	\multicolumn{21}{c}{}\\
B2111+46&  0.610& 12& 10&$-$1.1& 1.0&15.8&$-$10& 13&88.7& 1.3& 86.2& 1.3& 59.07& 0.02&  683&  483&  647&456& 318&  215\\
        &  0.925& 13& 25&$-$1.4& 2.8& 3.1& $-$9& 25&82.1& 1.2& 79.1& 1.1& 55.29& 0.04&  591&  414&  551&384& 281&  189\\
        &  1.408& 13&  4&$-$1.4& 0.5&61.2& $-$9&  4&86.9& 1.9& 82.8& 1.8& 51.12& 0.02&  658&  464&  601&422& 243&  162\\
        &  1.642& 14& 28&$-$1.5& 3.0& 1.8& $-$9& 26&75.7& 0.5& 74.5& 0.4& 49.13& 0.06&  507&  352&  492&341& 225&  149\\
        &  4.850& 16& 42&$-$2.5& 6.6& 1.1& $-$6& 22&75.1& 0.6& 73.2& 0.6& 46.07& 0.03&  499&  346&  476&330& 200&  132\\
\noalign{\smallskip} 
\hline
\hline
\noalign{\smallskip} 
\end{tabular}
\end{table*}

\begin{table*}
\caption{The table lists measurements of $\Delta \phi$ and the delay height $\rm r_{\rm delay}^{\rm BCW}$
estimates for the BCW methods corresponding to OTS and PPCS measurements. The geometrical height 
estimate of Kijak \& Gil (2003) is given as $\rm r_{\rm geo}^{\rm KG}$. The parameter 
$s=(\frac{\rm r_{\rm geo}^{\rm OTS}}{2\rm r_{\rm delay}^{\rm BCW}})^{0.5}$ is also listed (except the ones
with negative $\rm r_{\rm delay}^{\rm BCW}$ values). See
text for further details. \label{tab3}} 
\begin{tabular}{c@{\hspace{0.15cm}}
                l@{\hspace{0.15cm}}r@{$\pm$}
		l@{\hspace{0.15cm}}r@{$\pm$}
		l@{\hspace{0.15cm}}r@{$\pm$}
		l@{\hspace{0.15cm}}r@{$\pm$}
		l@{\hspace{0.15cm}}
		c@{\hspace{0.15cm}}
		c@{\hspace{0.15cm}}
		l@{\hspace{0.15cm}}l}
\noalign{\smallskip} 
\hline
\hline
\noalign{\smallskip} 
PSR B&
Freq&
\multicolumn{2}{c}{$\Delta\phi_{\rm OTS}$}&
\multicolumn{2}{c}{$\Delta\phi_{\rm PPCS}$}&
\multicolumn{2}{c}{$\rm r_{\rm delay}^{\rm BCW}(OTS)$}&
\multicolumn{2}{c}{$\rm r_{\rm delay}^{\rm BCW}(PPCS)$}&
${\rm r_{\rm geo}^{\rm KG}}$&
$\frac{(\rm r_{\rm geo}^{\rm OTS})}{(\rm  r_{\rm delay}^{\rm BCW} (OTS))}$&
$s$\\
     &
(GHz)&
\multicolumn{2}{c}{($^\circ$)}&
\multicolumn{2}{c}{($^\circ$)}&   
\multicolumn{2}{c}{(km)}&
\multicolumn{2}{c}{(km)}&
(km)& 		
    &
    &\\
\hline

B0301+19   &   0.408  &  -0.48  &    0.54  &   -1.18   &   0.24  &  139  &  157   & 341  &   70  &  567   &   0.87  & 0.66   \\
   &   0.610  &   0.01  &    0.68  &   -1.09   &   0.05  &   -4  &  198   & 315  &   15  &  511   & -42.71  & $--$       \\
   &   0.925  &   0.16  &    0.35  &   -0.28   &   0.13  &  -47  &  101   &  81  &   40  &  458   &  -2.64  & $--$       \\
   &   1.408  &  -0.25  &    0.32  &   -0.22   &   0.08  &   72  &   92   &  64  &   24  &  411   &   1.71  & 0.92   \\
   &   1.642  &  -0.50  &    0.88  &   -0.68   &   0.66  &  146  &  254   & 197  &  192  &  395   &   0.55  & 0.52   \\
	\multicolumn{13}{c}{}\\
B0525+21   &   0.408  &   0.14  &    0.29  &    0.28   &   0.12  & -115  &  232   &-219  &  100  &  971   & -16.78  & $--$       \\
   &   0.610  &  -0.30  &    0.49  &   -0.48   &   0.05  &  238  &  386   & 378  &   40  &  875   &   9.51  & 2.18   \\
   &   0.925  &  -0.21  &    0.36  &   -0.28   &   0.08  &  165  &  286   & 224  &   68  &  785   &  11.52  & 2.40   \\
   &   1.408  &   0.09  &    0.29  &    0.18   &   0.05  &  -73  &  228   &-140  &   45  &  704   & -27.89  & $--$       \\
   &   1.642  &   0.29  &    0.49  &   -0.11   &   0.13  & -232  &  389   &  87  &  106  &  676   &  -8.41  & $--$       \\
   &   4.850  &  -0.50  &    0.29  &   -0.66   &   0.19  &  392  &  234   & 516  &  150  &  510   &   3.60  & 1.34   \\
	\multicolumn{13}{c}{}\\
B1039-19   &   0.610  &  -0.89  &    0.39  &   -1.49   &   0.09  &  258  &  113   & 432  &   28  &  499   &   2.54  & 1.12   \\
   &   0.925  &  -1.40  &    0.53  &   -2.19   &   0.43  &  407  &  153   & 634  &  126  &  448   &   1.30  & 0.80   \\
   &   1.408  &  -0.01  &    0.28  &   -0.97   &   0.09  &    5  &   81   & 280  &   27  &  402   & 101.80  & 7.13   \\
   &   1.642  &   0.25  &    0.54  &   -0.90   &   0.40  &  -74  &  158   & 261  &  116  &  386   &  -6.69  & $--$       \\
	\multicolumn{13}{c}{}\\
B1737+13   &   0.408  &  -1.76  &    0.88  &   -4.00   &   1.19  &  295  &  148   & \multicolumn{2}{c}{--}      &  485   &   1.21  & 0.78   \\
   &   0.610  &  -3.07  &    1.63  &   -4.10   &   0.90  &  513  &  273   & \multicolumn{2}{c}{--}      &  437   &   1.10  & 0.74   \\
   &   0.910  &  -1.11  &    0.38  &   -0.18   &   0.27  &  185  &   64   &  30  &   46  &  394   &   1.73  & 0.93   \\
   &   1.408  &  -1.49  &    0.46  &   -0.24   &   0.78  &  249  &   78   &  40  &  131  &  351   &   1.79  & 0.94   \\
   &   1.642  &  -1.01  &    0.27  &   -0.00   &   0.82  &  170  &   46   &   0  &  138  &  337   &   2.15  & 1.03   \\
	\multicolumn{13}{c}{}\\
B2045-16   &   0.408  &   0.02  &    0.37  &    0.11   &   0.03  &  -11  &  152   & -47  &   13  &  730   & -70.41  & $--$       \\
   &   0.606  &  -0.38  &    0.80  &   -0.14   &   0.00  &  155  &  328   &  57  &    1  &  659   &   7.12  & 1.88   \\
   &   0.925  &  -0.34  &    0.44  &   -0.09   &   0.10  &  139  &  183   &  39  &   43  &  590   &   4.92  & 1.56   \\
   &   1.408  &  -0.55  &    0.23  &   -0.27   &   0.01  &  228  &   94   & 113  &    6  &  529   &   3.62  & 1.34   \\
   &   1.642  &  -0.95  &    0.29  &   -0.56   &   0.02  &  391  &  121   & 231  &   10  &  508   &   1.94  & 0.98   \\
   &   4.850  &  -0.65  &    0.20  &   -0.50   &   0.18  &  266  &   83   & 206  &   74  &  384   &   1.76  & 0.94   \\
	\multicolumn{13}{c}{}\\
B2111+46   &   0.610  &  -2.65  &    1.37  &   -2.21   &   0.06  &  560  &  291   & 467  &   13  &  446   &   1.21  & 0.78   \\
   &   0.925  &  -1.62  &    1.23  &   -1.10   &   0.16  &  344  &  260   & 234  &   35  &  400   &   1.71  & 0.92   \\
   &   1.408  &  -1.08  &    1.96  &   -0.77   &   0.03  &  228  &  414   & 163  &    7  &  358   &   2.87  & 1.19   \\
   &   1.642  &  -1.87  &    0.55  &   -1.12   &   0.22  &  396  &  116   & 237  &   46  &  344   &   1.27  & 0.79   \\
   &   4.850  &  -2.73  &    0.73  &   -3.62   &   0.27  &  578  &  156   & 765  &   58  &  260   &   0.86  & 0.65   \\

\noalign{\smallskip} 
\hline
\hline
\noalign{\smallskip} 
\end{tabular}
\end{table*}

\begin{table*}
\caption{The table lists measurements of $\Delta \phi$ and the delay height $\rm r_{\rm delay}^{\rm GGa}$
estimates for the BCW methods corresponding to OTS and PPCS measurements. The geometrical height 
estimate of Kijak \& Gil (2003) is given as $\rm r_{\rm geo}^{\rm KG}$. The parameter 
$s=(\frac{\rm r_{\rm geo}^{\rm OTS}}{2\rm r_{\rm delay}^{\rm GGa}})^{0.5}$ is also listed. See
text for further details.\label{tab4}} 
\begin{tabular}{c@{\hspace{0.15cm}}
                l@{\hspace{0.15cm}}r@{$\pm$}
		l@{\hspace{0.15cm}}r@{$\pm$}
		l@{\hspace{0.15cm}}r@{$\pm$}
		l@{\hspace{0.15cm}}r@{$\pm$}
		l@{\hspace{0.15cm}}
		c@{\hspace{0.15cm}}
		c@{\hspace{0.15cm}}
		l@{\hspace{0.15cm}}l}
\noalign{\smallskip} 
\hline
\hline
\noalign{\smallskip} 
PSR B&
Freq&
\multicolumn{2}{c}{$\Delta\phi_{\rm OTS}$}&
\multicolumn{2}{c}{$\Delta\phi_{\rm PPCS}$}&
\multicolumn{2}{c}{$\rm r_{delay}^{GGa}(OTS)$}&
\multicolumn{2}{c}{$\rm r_{delay}^{GGa}(PPCS)$}&
${\rm r_{\rm geo}^{\rm KG}}$&
$\frac{(\rm r_{\rm geo}^{\rm OTS})}{(\rm r_{\rm delay}^{\rm GGa} (OTS))}$&
$s$\\
     &
(GHz)&
\multicolumn{2}{c}{($^\circ$)}&
\multicolumn{2}{c}{($^\circ$)}&   
\multicolumn{2}{c}{(km)}&
\multicolumn{2}{c}{(km)}&
(km)& 		
    &
    &\\
\hline

B2045-16  &    0.408  &  -2.41   &   0.35  &   -2.32   &   0.03 &  1970 &   287  & 1898  &   25  &  730   &   0.41   &   0.64 \\
  &    0.606  &  -2.29   &   0.36  &   -2.20   &   0.00 &  1873 &   298  & 1803  &    3  &  659   &   0.57   &   0.75 \\
  &    0.925  &  -2.31   &   0.46  &   -2.06   &   0.10 &  1890 &   382  & 1689  &   87  &  590   &   0.36   &   0.60 \\
  &    1.408  &  -2.21   &   0.24  &   -1.94   &   0.01 &  1814 &   201  & 1592  &   13  &  529   &   0.45   &   0.67 \\
  &    1.642  &  -2.47   &   0.37  &   -2.08   &   0.02 &  2026 &   305  & 1704  &   21  &  508   &   0.37   &   0.61 \\
	\multicolumn{13}{c}{}\\

B2111+46   &   0.610  &  -3.31  &    1.30  &   -2.97   &   0.06  & 1403  &  549  & 1259  &   27  &  446   &   0.49   &   0.70\\
   &   0.925  &  -2.61  &    1.24  &   -2.10   &   0.16  & 1105  &  525  &  891  &   71  &  400   &   0.53   &   0.76\\
   &   1.408  &  -2.60  &    1.42  &   -2.47   &   0.03  & 1100  &  601  & 1045  &   14  &  358   &   0.60   &   0.78\\
   &   1.642  &  -3.57  &    0.55  &   -2.82   &   0.22  & 1512  &  235  & 1195  &   94  &  344   &   0.33   &   0.57\\
   &   4.850  &  -3.12  &    1.29  &   -3.35   &   0.27  & 1322  &  545  & 1417  &  117  &  260   &   0.39   &   0.62\\

\noalign{\smallskip} 
\hline
\hline
\noalign{\smallskip} 
\end{tabular}
\end{table*}
\end{document}